\documentclass[showpacs,showkeys,amsmath,amssymb,12pt]{revtex4}

\usepackage{graphicx}

\begin{document}


\title{Analytical Kerr--Sen Dilaton-Axion Black Hole Lensing in the Weak Deflection Limit}
\author{Galin N. Gyulchev\footnote{E-mail: gyulchev@phys.uni-sofia.bg}, Stoytcho S. Yazadjiev\footnote{E-mail: yazad@phys.uni-sofia.bg} }
\affiliation{Department of Theoretical Physics, Faculty of Physics, Sofia University,\\
5 James Bourchier Boulevard, 1164 Sofia, Bulgaria
}
\date{\today}


\begin{abstract}

We investigate analytically gravitational lensing by charged, stationary, axially symmetric Kerr--Sen dilaton-axion black hole in the weak deflection limit. Approximate solutions to the lightlike equations of motion are present up to and including third-order terms in $M/b$, $a/b$ and $r_{\alpha}/b$, where $M$ is the black hole mass, $a$ is the angular momentum, $r_{\alpha}=Q^2/M$, $Q$ being the charge and $b$ is the impact parameter of the light ray. We compute the positions of the two weak field images, the corresponding signed and absolute magnifications up to post-Newtonian order. It is shown that there are static post-Newtonian corrections to the signed magnification and their sum as well as to the critical curves, which are functions of the charge. The shift of the critical curves as a function of the lens angular momentum is found, and it is shown that they decrease slightly with the increase of the charge. The point-like caustics drift away from the optical axis and do not depend on the charge. All of the lensing quantities are compared to particular cases as Schwarzschild and Kerr black holes as well as the Gibbons--Maeda--Garfinkle--Horowitz--Strominger black hole.

\end{abstract}


\maketitle


\section{Introduction}

The gravitational lensing can be used for a confirmation of the generalized gravity theories. One of the most promising avenue realizing the unified theories is the string theory, which in the low-energy limit reduces to the Einstein--Maxwell dilaton-axion gravity. Nowadays, it is widely believed that in nearly every galaxy is hosted a massive black hole. Inspired by this hypothesis, we model the massive dark object in the center of the Milky way Galaxy as a Kerr--Sen black hole dilaton-axion generalization of the classical Kerr solution with the aim of investigating the charge and the dilaton-axion field effect over the lensing oservables.

The problem for the bending of light rays has developed in two main directions, namely a gravitational lensing in the strong and the weak deflection limit. The former considers the multiple winding of the photons moving close the lens before reaching the observer. As a result, an infinite series of highly demagnified images around either side of the lens shadow appear. The latter studies the moving of photons in the asymptotically flat spacetime on a distance much larger than the lens gravitational radius. In this case the deflection of the light ray has small values and two weak field images on each side of the lens arise. There exist intermediate case of consideration, which bridge the gap between the above two limiting cases.

In the weak deflection limit, the theory of gravitational lensing has been described successfully with perturbative methods applied to the theory of gravity under consideration. The first efforts in this direction have been applied for a Schwarzschild point-mass lens \cite{SEF, PettersLevine}. The rotational case has been studied for the first time up to a post-Newtonian expansion by Epstein and Shapiro \cite{EpsteinShapiro} and after that by Richter and Matzner \cite{RichterMatzner}. Later on, Bray \cite{Bray} investigated the multi-imaging aspect of Kerr black hole lensing resolving the equations of motion for a light ray up to and including second order terms in the scaled black hole mass $m/r_{\rm min}$ and angular momentum $a/r_{\rm min}$, where $r_{\rm min}$ is the distance of closest approach. Gravitational lensing by rotating stars has been considered by Glinstein \cite{Glinstein} and later on by Sereno \cite{Sereno1}. The weak field Reissner-Nordstrom black hole lensing has been done by Sereno \cite{Sereno2}. Asada and Kasai \cite{AsadaKasai} have found that up to the first order in the gravitational constant $G$, a rotating lens is not distinguishable from a not-rotating one. They have found that because of the global translation of the center of lens mass the Kerr lens is observationally equivalent to the Schwarzschild one at linear order in the mass $m$ and the specific lens angular momentum $a$. Later on, Asada, Kasai, and Yamamoto have shown \cite{AsadaKasaiYamamoto}, that the nonlinear coupling breaks the degeneracy so that the rotational effect becomes in principle separable for multiple images or a single source. After that, Sereno \cite{Sereno3} considered gravitational lensing in the metric theories of gravity in post-post-Newtonian order with gravitomagnetic field. Keeton and Petters \cite{KeetonPetters} have made a step forward and developed a general formalism for lensing by spherically symmetric lenses up to post-post-Newtonian order. Sereno and De Luca \cite{SerenoLuca} extended their approach to the case of Kerr black hole. Finally, Werner and Petters \cite{WernerPetters} applied a simpler method based on the analysis by Asada, Kasai, and Yamamoto to derive image positions and magnifications up to post-Newtonian order. Using the degeneracy in the case of Kerr black hole lensing in the weak deflection limit, they presented lensing observables for the two weak field images in post-Newtonian terms. After that Vibhadra and Keeton \cite{VirbhadraKeeton} have examined numerically the time delay and magnification centroid due to gravitational lensisng by black holes and naked singularities.

Light deflection in a strong field regime received a wide attention after works by Darwin \cite{Darwin}. As a consequence, in the field of study a wave of a profound examinations arose. Many authors have discussed gravitational lensing in the strong deflection limit approximation in various metrics from general relativity, string theory and braneworld gravity \cite{LON, Viergutz, RauchBlandford, V1, Frittelli, Perlick1, Bozza1, Eiroa, Bhadra, SBS, Majumdar, Konoplya, Petters, Bozza2, G3, G4, V2, KentaMaeda, ChenJing, AlievTalazan}. Amore and Arceo \cite{AmoreArceo} as well as Iyer and Petters \cite{IyerPetters} have presented a procedures interpolating between the weak and the strong deflection limits analysis. Besides, Perlick \cite{Perlick2} has considered lensing in the spherically symmetric case based on the light-like geodesics equations without approximations.

The purpose of this paper is to consider the gravitational lensing by rotating, dilaton-axion black holes and to explore how they differ from the Kerr and the static and spherically, symmetric Gibbons--Maeda--Garfinkle--Horowitz--Strominger black hole lensing previously studied in \cite{KeetonPetters}. Following the method of Sereno and De Luca \cite{SerenoLuca} in the present work we will concentrate our attention on the gravitational lensing in the weak deflection limit due to a stationary, axially symmetric Kerr--Sen dilaton-axion black hole in the heterotic string theory with the aim of investigating the influence of both the angular momentum and the charge on the behaviour of the bending angle, on the position of the images and on their magnifications as well as on the critical curves and caustics.

The paper is organized as follows. The Sec. II, we review the Kerr--Sen solution, derive the first order differential system for null geodesics and find the minimal distance of closest approach of the photons. In Sec. III, the lens equations in the weak deflection limit are derived using the geodesics equations. In Sec. IV, the lens equations are solved with standard perturbative methods. In Sec. V, the critical curves and the caustic structure are considered.  In Sec. VI we derive the signed magnification relations for the two weak field images and the total magnification. A discussion of the results is given in Sec. VII. In Appendixes A and B the resolution methods of the radial and angular integrals appearing in the geodesics equations are presented. We use geometrized units, in which the gravitational constant $G=1$ and the speed of light in vacuum $c=1$, throughout this paper.


\section{Rotating Kerr-Sen dilaton-axion black hole and null geodesics}

We consider a stationary, axially symmetric solution of the heterotic string theory field equations \cite{Sen}. The line element of this solution in generalized Boyer-Lindquist coordinates $\{t, r, \vartheta, \phi\}$ is given by
\begin{eqnarray}
ds^{2}=&-&\left(1-\frac{2Mr}{\rho^{2}}\right)dt^{2}+\rho^{2}\left(\frac{dr^{2}}{\Delta}+d\vartheta^{2}\right)-\frac{4Mra\sin^{2}{\vartheta}}{\rho^{2}}dt{d\phi} \nonumber  \\  &+&\left(r(r+r_{\alpha})+a^{2}+\frac{2Mra^{2}\sin^{2}{\vartheta}}{\rho^{2}} \right)\sin^{2}{\vartheta}d\phi^{2} \label{LineElement},
\end{eqnarray}
where
\begin{eqnarray}
    \Delta &=& r(r+r_{\alpha})-2Mr+a^{2}, \\
    \rho^{2} &=& r(r+r_{\alpha})+a^{2}\cos^{2}{\vartheta}.
\end{eqnarray}
Here $M$ is the mass of the black hole, $a=L/M$ is the angular momentum of the black hole in units of mass and $r_{\alpha}=Q^2/M$, where $Q$ is the charge of the black hole. In the particular case of a static black hole, \emph{i.e.} $a=0$ the solution (\ref{LineElement}) coincides with the Gibbons--Maeda--Garfinkle--Horowitz--Strominger (GMGHS) solution investigated in WDL regime from Keeton and Petters in \cite{KeetonPetters}. The particular case $r_{\alpha}=0$ reconstructs the Kerr solution investigated in WDL from Sereno and De Luca in \cite{SerenoLuca}.

The Kerr-Sen space is characterized by a spherical event horizon, which is the biggest root of the equation $\Delta=0$ and is equal to
\begin{eqnarray}
  r_{H}=\frac{2M-r_{\alpha}+\sqrt{(2M-r_{\alpha})^{2}-4a^{2}}}{2}.
\end{eqnarray}

\begin{figure}
  \includegraphics[width=12cm]{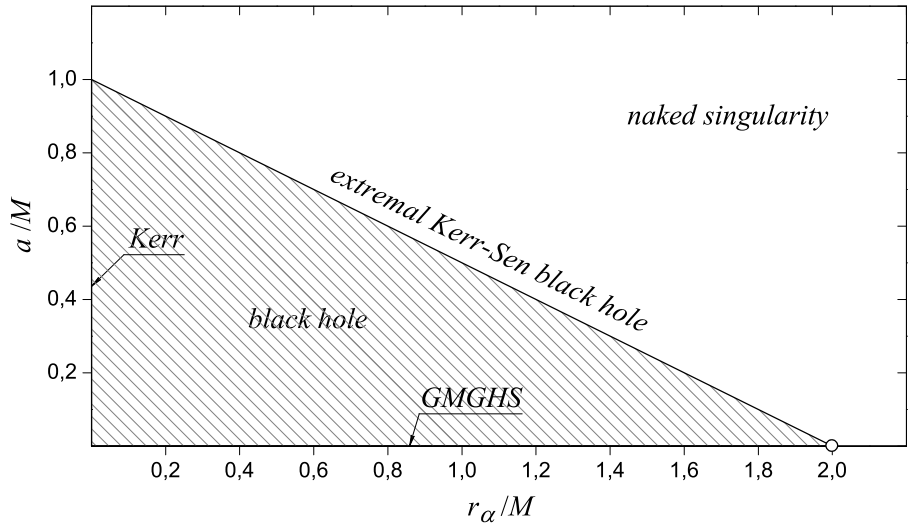}\\
  \caption{ Kerr--Sen, Gibbons--Maeda--Garfinkle--Horowitz--Strominger and Kerr black hole relations in the parametric space $\{a, r_{\alpha}\}$. Extremal Kerr--Sen black hole curve divides the space into black hole and naked singularity regions. The circle at point $(0, 2M)$ marks a naked singularity.}\label{BH-NS}
\end{figure}

Hence follows the regularity conditions of the event horizon $a<a_{cr}=\frac{1}{2}\Big{|}1-{\frac{r_{\alpha}}{2M}}\Big{|}$. Beyond this critical value of the spin an extremal Kerr--Sen black hole ($a=a_{cr}$) and a naked singularity appear ($a>a_{cr}$). The event horizon regularity condition in terms of the parameter $r_{\alpha}$ and the specific angular momentum $a$ is shown on Fig. \ref{BH-NS}. We will restrict our considerations up to and including critical angular momenta. The hypersurface on which the Killing vector $\frac{\partial}{\partial{t}}$ is isotropic, \textit{i.e} $g_{tt}=0$, is the so called ergosphere described by the equation
\begin{eqnarray}
  r_{\rm es}=\frac{2M-r_{\alpha}+\sqrt{(2M-r_{\alpha})^{2}-4a^{2}\cos^{2}\vartheta}}{2}.
\end{eqnarray}
In the case of gravitational lensing in the weak deflection limit regime the light ray has a minimal distance of a closest approach $r_{\rm min}>>r_{\rm es}$, so that the photons do not passing through the ergosphere.

The equations of motion for a ray of light can be derived via the Hamilton-Jacobi equations for metric (\ref{LineElement}), (see \cite{BlagaBlaga}). The lightlike fourth-order differential system is
\begin{eqnarray}
    \rho^{2}\dot{r}&=&\pm\sqrt{R(r)}, \label{Sys1} \\
    \rho^{2}\dot{\vartheta}&=&\pm\sqrt{\Theta(\vartheta)}, \label{Sys2} \\
    \rho^{2}\dot{\phi}&=&-aE+L_{z}\sin^{-2}\vartheta+\frac{a}{\Delta}[(r(r+r_{\alpha})+a^{2})E-aL_{z}], \label{Sys3} \\
    \rho^{2}\dot{t}&=&\frac{E(r(r+r_{\alpha})+a^{2})^{2}-2MraL_{z}}{\Delta}-a^{2}E\sin^{2}\vartheta, \label{Sys4}
\end{eqnarray}
where
\begin{eqnarray}
  R(r)&=&[aL_{z}-(r(r+r_{\alpha})+a^{2})E]^{2}-\Delta[(L_{z}-aE)^{2}+{\cal K}], \label{Func_R} \\
  \Theta(\vartheta)&=&{\cal K}-\cot^{2}\vartheta[L^{2}_{z}-\sin^{2}\vartheta{E}^{2}a^{2}]. \label{Func_Theta}
\end{eqnarray}

In Eqs. (\ref{Sys1}-\ref{Sys4}), the dot indicates the derivative with respect to the geodesic affine parameter, in Eqs. (\ref{Func_R}) and (\ref{Func_Theta}), ${\cal K}$ is a separation constant of motion, $E=-p_{t}$ is the energy at infinity, $L_{z}=p_{\phi}$ is the angular momentum with respect to the rotation axis of the black hole, while $a=L_{z}/M$ is the angular momentum per unit of mass.

The final expressions of the photon trajectory follows from Eqs. (\ref{Sys1}-\ref{Sys4}) in terms of integrals
\begin{eqnarray}\label{IntEqMotion1}
  \int^{r}{\frac{dr}{\pm\sqrt{R(r)}}}=\int^{\vartheta}{\frac{d\vartheta}{\pm\sqrt{\Theta(\vartheta)}}},
\end{eqnarray}
\begin{eqnarray}\label{IntEqMotion2}
    \triangle\phi=a\int^{r}{\frac{[(r(r+r_{\alpha})+a^{2})E-aL_{z}]}{\pm\Delta\sqrt{R(r)}}}dr+
                \int^{\theta}\frac{[L_{z}\sin^{-2}\theta-aE]}{\pm\sqrt{\Theta(\theta)}}d\theta.
\end{eqnarray}

The sign in front of $\sqrt{R(r)}$ and $\sqrt{\Theta(\vartheta)}$ is positive when the lower integration limit is smaller than the upper limit, and negative otherwise.

We consider a static source $\{r_{s}, \vartheta_{s}, \phi_{s}\}$ and a static observer $\{r_{o}, \vartheta_{o}, \phi_{o}\}$ located in the asymptotically flat region of the space-time, far away from the lens, so that $r_{s/o}\gg r_{\rm es}$. Taking into account the zero of the azimuth, we set $\phi_{o}=0$. Hereafter, we will also use the modified polar coordinate $\mu=\cos{\vartheta}$. Along the motion of the photon from the source to the observer the $r$ coordinate first decreases from $r_{s}$ to a minimum distance $r_{\rm min}$ and grows finally to $r_{o}$ approaching to the observer. Depending on the direction taken by the light ray at $\vartheta_{s}$, the photon can attain a maximum $\vartheta_{\rm max}$ or a minimum $\vartheta_{\rm min}$. If $\vartheta$ is initially growing ($\mu$ decreasing) the photon reaches a maximum angle $\vartheta_{\rm max}$ (a minimum $\mu$) and after that decreases reaching to the observer at $\vartheta_{o}$ ($\mu$ increases to get to $\mu_{o}$), otherwise for an initially decreasing. In the other case, if $\vartheta$ is initially decreasing ($\mu$ increasing) the photon reaches a minimum angle $\vartheta_{\rm min}$ (a maximum $\mu$) and after that increases getting to the observer at $\vartheta_{o}$ ($\mu$ decreases to get to $\mu_{o}$). The sign in front of $\sqrt{R(r)}$ is chosen to be positive if we are integrating from $r_{\rm min}$ to $r_{o}$ or $r_{s}$ and negative otherwise. By analogy the sign in front of $\sqrt{\Theta(\vartheta)}$ is chosen to be positive (or negative) if integration is from $\vartheta_{\rm min}$ (or $\vartheta_{\rm max}$) to $\vartheta_{o}$ or $\vartheta_{s}$.

In the weak deflection limit we can describe the light ray trajectory with the assumption that at infinity, where the black hole attraction is insignificant, it is a straight line. The approximate light ray can be identified by the photon angular momentum $\cal{J}$ and the separation constant of motion $\cal{K}$. For observer with coordinates $(r_{o},\vartheta_{o})$ in the Boyer-Lindquist system, we can describe the images with two celestial coordinates $\xi_{1}$ and $\xi_{2}$. By definition $\xi_{1}$ and $\xi_{2}$ are the coordinates of the image position projected along the tangent to the light ray at the observer into the plane through the black hole and normal to the line connecting it with the observer. These are observable quantities and are related to the constants of the motion $\cal{J}$ and $\cal{K}$ as can be shown. Using Eqs.(\ref{Sys1}-\ref{Sys3}) and further allowing $E=1$ and $r_{o}\rightarrow\infty$, we get
\begin{eqnarray}
  \xi_{1}=r_{o}^{2}\sin{\theta_{o}}\frac{d\phi}{dr}\bigg|_{r_{o}\rightarrow\infty}&=&-\frac{ {\cal{J}} }{ \sqrt{1-\mu_{o}^2} }, \label{Observables1}  \\
  \xi_{2}=r_{o}^{2}\frac{d\theta}{dr}\bigg|_{r_{o}\rightarrow\infty}&=&\pm\sqrt{ {\cal{K}} + a^2\mu_{o}^2 - {\cal{J}}^2\frac{\mu_{o}^2}{1-\mu_{o}^2} }. \label{Observables2}
\end{eqnarray}

The roots of $R(r)$ define the inversion point in the radial motion. For weak gravitational lensing there is only one inversion point, which represents the closest approach distance, the largest of which is $r_{\rm min}$. For equatorial observer $\vartheta_{o}=\pi/2$ and large $r_{o}$ the observable distance of an image with respect to the black hole is $b\simeq\sqrt{\xi_{1}^2+\xi_{2}^2}=\sqrt{{\cal{J}}^2+{\cal{K}}}$, which is exactly the spherically symmetric impact parameter. The impact parameter is a conserved quantity defined geometrically by the perpendicular distance from the center of the lens to the tangent of the light ray at the observer. We expect for small deflections $r_{\rm min}$ to be of the same order. Therefore, following \cite{Bray, SerenoLuca} for the solution we suggest that
\begin{equation}
    r_{\rm min} \simeq \sqrt{{\cal{J}}^2+{\cal{K}}}\left\{1+\sum_{i,j,k=0}^{4}c_{mar_{\alpha}}\epsilon_{m}^{i}\epsilon_{a}^{j}\epsilon_{r_{\alpha}}^{k}\right\},
\end{equation}
where we are introducing three independent expansion parameters in terms of the invariants of motion
\begin{eqnarray}
    \epsilon_{m} &=& \frac{M}{\sqrt{{\cal{J}}^2+{\cal{K}}}}, \\
    \epsilon_{a} &=& \frac{a}{\sqrt{{\cal{J}}^2+{\cal{K}}}}, \\
    \epsilon_{r_{\alpha}} &=& \frac{r_{\alpha}}{\sqrt{{\cal{J}}^2+{\cal{K}}}}.
\end{eqnarray}
Here the coefficients $c_{mar_{\alpha}}$ are real numbers, and the summation is over all possible combinations of epsilon powers $i, j$ and $k$ up to and including fourth order terms.

Solving the equation $R(r_{\rm min})=0$ we find for the closest approach distance
\begin{eqnarray}\label{CAD}
    r_{\rm min} &\simeq& {({\cal{J}}^2+\cal{K})}^{1/2} \bigg{\{} 1-\frac{M}{\sqrt{{\cal{J}}^2+\cal{K}}}-\frac{3M^2}{2({\cal{J}}^2+\cal{K})} - \frac{4M^3}{{({\cal{J}}^2+\cal{K})}^{3/2}} - \frac{105M^4}{8{({\cal{J}}^2+\cal{K})}^{2}}  \\
    &-& \frac{J^2a^2}{2{({\cal{J}}^2+\cal{K})}^2} - \frac{{\cal{J}}^2({\cal{J}}^2-4{\cal{K}})a^4}{8{({\cal{J}}^2+\cal{K})}^{4}} - \frac{r_{\alpha}}{2\sqrt{({\cal{J}}^2+\cal{K})}} + \frac{r_{\alpha}^2}{8{({\cal{J}}^2+\cal{K})}} - \frac{r_{\alpha}^4}{128{({\cal{J}}^2+\cal{K})}^{2}} \nonumber \\
    &+& \frac{2JMa}{{({\cal{J}}^2+\cal{K})}^{3/2}}  +  \frac{6JM^2a}{{({\cal{J}}^2+\cal{K})}^{2}} - \frac{2J^2Ma^2}{{({\cal{J}}^2+\cal{K})}^{5/2}} + \frac{2{\cal{J}}({\cal{J}}^2-{\cal{K}})Ma^3}{{({\cal{J}}^2+\cal{K})}^{7/2}} +  \frac{24JM^3a}{{({\cal{J}}^2+\cal{K})}^{3/2}} \nonumber \\
    &+& \frac{(8{\cal{K}}-51{\cal{J}}^2)M^2a^2}{4{({\cal{J}}^2+\cal{K})}^{3}} - \frac{{\cal{J}}Mar_{\alpha}}{{({\cal{J}}^2+\cal{K})}^{3}} + \frac{Mr_{\alpha}}{2{({\cal{J}}^2+\cal{K})}} + \frac{2M^2r_{\alpha}}{{({\cal{J}}^2+\cal{K})}^{3/2}} + \frac{35M^3r_{\alpha}}{4{({\cal{J}}^2+\cal{K})}^{2}} \nonumber \\
    &-& \frac{Mr_{\alpha}^3}{16{({\cal{J}}^2+\cal{K})}^{2}} - \frac{15M^2r_{\alpha}^2}{16{({\cal{J}}^2+\cal{K})}^{2}} + \frac{{\cal{J}}^2a^2r_{\alpha}^2}{16{({\cal{J}}^2+\cal{K})}^{3}} - \frac{8{\cal{J}}M^2ar_{\alpha}}{{({\cal{J}}^2+\cal{K})}^{5/2}} + \frac{5{\cal{J}}^2Ma^2r_{\alpha}}{4{({\cal{J}}^2+\cal{K})}^{3}} + \mathcal{O}(\epsilon^5) \bigg{\}}. \nonumber
\end{eqnarray}
An expression for the minimal distance including terms $\mathcal{O}(\epsilon^4)$ can be found in \cite{SerenoLuca} for the case of Kerr black hole $(r_{\alpha}=0)$. Up to the given formal order in the expansion parameter the expression (\ref{CAD}) coincides with the minimal distance of closest approach calculated in \cite{KeetonPetters} for the static case of the Gibbons--Maeda--Garfinkle--Horowitz--Strominger black hole $(a=0)$.


\section{Lens Equations}

Let us consider the lens equations. The light ray trajectory can be described with the integrals of motion ${\cal{J}}$ and ${\cal{K}}$. Therefore, once we have connected the image positions with the integrals of motion and solved the geodesic equations, Eqs. (\ref{IntEqMotion1}) and (\ref{IntEqMotion2}) by $\mu_{s}$ and $\phi_{s}$ we can find the lens mapping $\{\mu_{s},\phi_{s}\}\mapsto\{\theta_{1},\theta_{2}\}$, which connect the source and the image positions. Following the papers \cite{Bray} and \cite{SerenoLuca}, which have developed the Kerr black hole lensing in the weak deflection limit, we construct the Kerr--Sen black hole lens equations. We find approximate analytical solutions of the isotropic equations of motion, which are corrected by small parameters $M/b$, $a/b$ and $r_{\alpha}/b$, where $M$ is a black hole mass, $a$ is the angular momentum and $r_{\alpha}=Q^2/M$ is the square of the charge per unit of mass. The methods for the resolution of the radial and angular integrals are described in the Appendices A and B, respectively. The geodesic equation, Eq. (\ref{IntEqMotion1}) representing the polar motion of the photon reduces to
\begin{equation}\label{PolarLensEq}
    \mu_{s}=-\mu_{o}\cos{\delta}+(-1)^{k}\sin{\delta}\left( \frac{ {\cal{K}} }{ {\cal{J}}^2+{\cal{K}} }-\mu_{o}^2 \right)^{1/2},
\end{equation}
where
\begin{eqnarray}\label{delta}
  \delta &=& \frac{4M}{ \sqrt{ {\cal{J}}^2+{\cal{K}} } } + \frac{15\pi M^2}{ 4( {\cal{J}}^2+{\cal{K}} ) } - \frac{8{\cal{J}}Ma}{ ( {\cal{J}}^2+{\cal{K}} )^{3/2} }
  + \frac{128{\cal{J}}M^3}{ 3( {\cal{J}}^2+{\cal{K}} )^{3/2} } - \frac{15\pi{\cal{J}}M^2a}{ ( {\cal{J}}^2+{\cal{K}} )^{2} }  \nonumber \\
  &+& 2\left[ \mu_{o}^2+\frac{5{\cal{J}}^2-3{\cal{K}}}{ {\cal{J}}^2+{\cal{K}} } + \frac{ {\cal{K}} }{ \frac{ {\cal{J}}^2\mu_{o}^2 }{\mu_{o}^2-1} + {\cal{K}} } \right]\frac{ Ma^2 }{ ( {\cal{J}}^2+{\cal{K}} )^{3/2} }
  - \frac{\pi r_{\alpha}^2}{ 16( {\cal{J}}^2+{\cal{K}} ) } - \frac{3\pi M r_{\alpha}}{ 4( {\cal{J}}^2+{\cal{K}} ) }   \nonumber \\
  &-& \frac{16M^2r_{\alpha}}{ ( {\cal{J}}^2+{\cal{K}} )^{3/2} } + \frac{3\pi{\cal{J}}Mar_{\alpha}}{ 2( {\cal{J}}^2+{\cal{K}} )^{2} } - \frac{ ({\cal{J}}^2+{\cal{K}})^{3/2} }{6}\frac{r_{s}^3+r_{o}^3}{r_{s}^3r_{o}^3} - ( {\cal{J}}^2+{\cal{K}} )^{1/2}\frac{r_{s}+r_{o}}{r_{s}r_{o}} \nonumber \\
  &+& \frac{ \sqrt{ {\cal{J}}^2+{\cal{K}} }\mu_{o}^2(1-\mu_{o}^2) }{2[\mu_{o}^2{\cal{J}}^2-{\cal{K}}(1-\mu_{o}^2)]}\frac{r_{s}+r_{o}}{r_{s}r_{o}}a^2 + \mathcal{O}(\epsilon^4).
\end{eqnarray}
For the case of the Kerr black hole ($r_{\alpha}=0$) $\delta$ reduces to the result found in \cite{SerenoLuca}. It can be shown that $\delta$ describes the deflection angle of the light ray when we allow the source and the observer to be at infinity. In the static case ($a=0$), $\delta$ describes GMGHS black hole deflection angle. The parameter $k$ in the equation (\ref{PolarLensEq}) accounts for the kind of the photon trajectory. $k$ is even (odd), when $\vartheta$ is increasing (decreasing) function with maximum $\vartheta_{\rm max}$ (minimum $\vartheta_{\rm min}$) and the photons reached to the observer from below (above) the lens.

The geodesic equation Eq. (\ref{IntEqMotion2}) describes the azimuthal motion of the photon. Taking account Eq. (\ref{PolarLensEq}) the Eq. (\ref{IntEqMotion2}) reads
\begin{eqnarray}\label{AzimutalLensEq}
   -\phi_{s} &=& \frac{ {\cal{J}} }{ | {\cal{J}} | }\pi + \frac{ {\cal{J}}\delta }{ \sqrt{ {\cal{J}}^2+{\cal{K}} } }\frac{1}{1-\mu_{o}^2}\left[ 1-(-1)^k\delta\frac{\mu_{o}}{1-\mu_{o}^2}\sqrt{ \frac{ {\cal{K}} }{ {\cal{J}}^2+{\cal{K}} }-\mu_{o}^2 } \right]  \nonumber \\
   &+& \frac{4Ma}{ {\cal{J}}^2+{\cal{K}} } + \delta \phi_{s},
\end{eqnarray}
where $\delta\phi_{s}$ have a contribution of order $\sim\epsilon^3$,
\begin{eqnarray}\label{ALE_Corection}
    \delta \phi_{s} &=& \frac{ 5\pi M a }{ ({\cal{J}}^2+{\cal{K}})^{3/2} } - \frac{64{\cal{J}}[2({\cal{J}}^2+{\cal{K}})\mu_{o}^4+(2{\cal{J}}^2-{\cal{K}})\mu_{o}^2-{\cal{K}}]M^3}{ 3({\cal{J}}^2+{\cal{K}})^{3}(1-\mu_{o}^2)^{3} } \nonumber \\
    &-& 2\left[ \frac{{\cal{K}}}{{\cal{K}}-({\cal{J}}^2+{\cal{K}})\mu_{o}^2} + 3 \right]\frac{{\cal{J}}Ma^2}{ ({\cal{J}}+{\cal{K}})^2 } - \frac{\pi M a r_{\alpha}}{2({\cal{J}}^2+{\cal{K}})^{3/2}}  \nonumber \\
    &-& \frac{ {\cal{J}}\mu_{o}^2a^2 }{2( {\cal{J}}^2\mu_{o}^2-{\cal{K}}(1-\mu_{o}^2) )}\frac{ r_{s}+r_{o} }{r_{s}r_{o}} + \frac{ {\cal{J}}[2({\cal{J}}^2+{\cal{K}})\mu_{o}^4+(2{\cal{J}}^2-{\cal{K}})\mu_{o}^2-{\cal{K}}] }{({\cal{J}}^2+{\cal{K}})(1-\mu_{o}^2)^{3}} \nonumber \\
    &\times&\left[ \frac{16M^2}{ {\cal{J}}^2+{\cal{K}} } - 4M\frac{r_{s}+r_{o}}{r_{s}r_{o}} - \frac{1}{3}({\cal{J}}^2+{\cal{K}})\frac{(r_{s}+r_{o})^2}{r_{s}^2r_{o}^2} \right]\frac{r_{s}+r_{o}}{r_{s}r_{o}} + \mathcal{O}(\epsilon^4).
\end{eqnarray}
For the case of Kerr black hole ($r_{\alpha}=0$) the Eqs. (\ref{AzimutalLensEq}) and (\ref{ALE_Corection}) have been already evaluated in \cite{SerenoLuca}.

The lens equations are usually given in terms of the apparent angular positions of the image over the plane on the sky $\{\theta_{1}, \theta_{2}\}$ and of the angular positions of the source $\{B_{1},B_{2}\}$ in absence of the lens. As the source and the observer are distributed in the asymptotically flat region of the space-time, we can introduce right-handed Cartesian coordinate system with the origin centered at the lens, such that the $Ox_{3}$ axis is oriented along the lens optical axis connecting the observer plane and the lens plane and the $Ox_{2}$ axis belongs to the lens plane and orthogonal to the $Ox_{1}$ axis. The last one traces the projection of the black hole axis of symmetry over the lens plane. In this setting $\theta_{1}$ and $\theta_{2}$ mark the image apparent angular coordinates measured along the $Ox_{1}$ axis and the $Ox_{2}$ axis, respectively. Then, the apparent angular image positions are connected with the celestial coordinates $\xi_{1}$ and $\xi_{2}$, (see Eqs. (\ref{Observables1}, \ref{Observables2})), and therefore are linked to the invariants of motion through the relations
\begin{eqnarray}
  r_{o}\frac{\tan{\theta_{1}}}{\sqrt{1+\tan^{2}{\theta}}} &=& -\frac{ {\cal{J}} }{ \sqrt{1-\mu_{o}^2} } \label{Observables3} \\
  r_{o}\frac{\tan{\theta_{2}}}{\sqrt{1+\tan^{2}{\theta}}} &=& -(-1)^{k}\sqrt{ {\cal{K}} + a^2\mu_{o}^2 - {\cal{J}}^2\frac{\mu_{o}^2}{1-\mu_{o}^2} }, \label{Observables4}
\end{eqnarray}
where $\theta$ is the angular separation of the image from the black hole and $\tan^{2}\theta=\tan^{2}\theta_{1}+\tan^{2}\theta_{2}$. The sign before the square root can be represented with the sign of the image component $\theta_{2}$, according to (\ref{Observables2}). From Eq. (\ref{Observables3}) it is obvious that the prograde photons (${\cal{J}}>0, {\cal{K}}=0$) will produce images standing on the left-hand side of the optical axis ($\theta_{1}<0$), while retrograde photons (${\cal{J}}<0, {\cal{K}}=0$) will produce images standing on the right-hand side of the optical axis ($\theta_{1}>0$).

In such a way, we can present the source polar and azimuthal coordinates $\{\mu_{s},\phi_{s}\}$ in terms of the observed image coordinates as a function of
the Kerr-Sen black hole parameters $M$, $a$, $r_{\alpha}$. Furthermore, the source position can be written in terms of angular source coordinates at which the source
would be seen by the observer in absence of the lens, i.e. for $M=a=r_{\alpha}=0$. The lens equation then can be obtained by first expressing $\mu_{s}$ and $\phi_{s}$ as a function of either \{$\theta_{1},\theta_{2}\}$ or $\{B_{1},B_{2}\}$ and finally equating the corresponding expressions (see Refs. \cite{Sereno4, Sereno5}),
\begin{eqnarray}\label{LensEquations1}
    \phi_{s}(\vartheta_{1},\vartheta_{2}|M,a,r_{\alpha})&=&\phi_{s}(B_{1},B_{2}|M=0,a=0,r_{\alpha}=0), \label{LensEquations11} \\
    \mu_{s}(\vartheta_{1},\vartheta_{2}|M,a,r_{\alpha})&=&\mu_{s}(B_{1},B_{2}|M=0,a=0,r_{\alpha}=0). \label{LensEquations12}
\end{eqnarray}

Solving the system of Eqs. (\ref{LensEquations11}) and (\ref{LensEquations12}) we can find the source angular coordinates $\{B_{1},B_{2}\}$ as a function of angular position of the images. From the other side, taking advantage of the lensing geometry, we can find the angular coordinates of the source more easily and express them in terms of radial coordinates \cite{SerenoLuca},
\begin{eqnarray}
    D_{s}\tan{B_{1}} &=& r_{s}\sin{\phi_{s}}\sqrt{1-\mu_{s}^2}, \label{AngularSourceAngles1} \\
    D_{s}\tan{B_{2}} &=& r_{s}( \mu_{s}\sqrt{1-\mu_{o}^2}-\mu_{o}\sqrt{1-\mu_{s}^2}\cos{\phi_{s}} ). \label{AngularSourceAngles2}
\end{eqnarray}
Here $D_{s}$ is the angular diameter distance measured along the optical axis from the observer plane to the source plane. The angular diameter distances can be related with the radial coordinates in the following way
\begin{eqnarray}
  D_{d} &=& r_{o}, \label{AngularDiameterDistances1} \\
  D_{ds} &=& -r_{s}[\mu_{o}\mu_{s}+\cos{\phi_{s}}\sqrt{(1-\mu_{o}^2)(1-\mu_{s}^2)}], \label{AngularDiameterDistances2} \\
  D_{s} &=& D_{d}+D_{ds}, \label{AngularDiameterDistances3}
\end{eqnarray}
where $D_{d}$ and $D_{ds}$ are the angular diameter distances measured along the optical axis from the observer plane to the lens plane, and from the plane of the lens to the source plane, respectively.

Finally, we have all elements to construct the lens equations in the classical form. Plugging Eqs. (\ref{Observables3}) and (\ref{Observables4}) in Eqs. (\ref{PolarLensEq}) and (\ref{AzimutalLensEq}) and after that exploiting Eqs. (\ref{AngularSourceAngles1}) and (\ref{AngularSourceAngles2}) we get
\begin{eqnarray}\label{LensMapping}
  B_{1} &=& B_{1}(\theta_{1},\theta_{2}), \\
  B_{2} &=& B_{2}(\theta_{1},\theta_{2}).
\end{eqnarray}

Examining gravitational lensing in the weak deflection limit regime the Kerr--Sen lens equations can be written in a simple form if we introduce a series expansion parameter $\varepsilon$ based on the angular Einstein ring. In terms of radial coordinates, the Einstein angular radius for a static observer is defined as \cite{SerenoLuca}
\begin{equation}\label{EinsteinRadius}
    \theta_{\rm E}=\sqrt{4M\frac{r_{s}}{r_{o}(r_{o}+r_{s})}}.
\end{equation}
The expansion parameter is then defined as \cite{KeetonPetters, SerenoLuca}
\begin{equation}
    \varepsilon=\frac{\theta_{\rm E}}{4D},
\end{equation}
where $D=r_{s}/(r_{o}+r_{s})$. In a way similar to the resolution of the geodesic equations, we can perform the expansion in terms of three parameters, $\varepsilon_{M}\equiv \varepsilon$, $\varepsilon_{a}=(\frac{a}{M})\varepsilon$ and $\varepsilon_{r_{\alpha}}=(\frac{r_{\alpha}}{M})\varepsilon$. Terms of different physical order are then collected up to a given formal order in $\varepsilon$. Holding contributions in order of $\varepsilon$ we assume that the solution of the lens equations can be written as
\begin{eqnarray}
  \theta_{1} &=& \theta_{\rm E}(\theta_{1(0)}+\theta_{1(1)}\varepsilon+\mathcal{O}(\varepsilon^2)), \label{ImagePositions1} \\
  \theta_{2} &=& \theta_{\rm E}(\theta_{2(0)}+\theta_{2(1)}\varepsilon+\mathcal{O}(\varepsilon^2)). \label{ImagePositions2}
\end{eqnarray}
The angular separation of the image from the black hole can be represented as
\begin{eqnarray}
  \theta = \theta_{\rm E}(\theta_{(0)}+\theta_{(1)}\varepsilon+\mathcal{O}(\varepsilon^2)). \label{ImagePositions3}
\end{eqnarray}
Therefore, the coefficients $\theta_{i(j)}$ and $\theta_{(j)}$ are polynomial of $j$-th order in $(a/M)^{m}(r_{\alpha}/M)^{n}$, where $m+n\leq j$. The source position also can be rescaled as $B_{i}=\theta_{\rm E}\beta$. Under the assumption for small angles (thin lens approximation) it can be shown that up to the given order in the expansion parameter $\varepsilon$ the lens equations acquire the expected form
\begin{eqnarray}
  B_{1} &=& \theta_{1} - D\hat{\alpha}_{1}(\theta_{1},\theta_{2}) \label{WDLLensEq1} \\
  B_{2} &=& \theta_{2} - D\hat{\alpha}_{2}(\theta_{1},\theta_{2}) \label{WDLLensEq2},
\end{eqnarray}
where $\hat{\alpha}$ is the angle of deflection of the light ray defined as the angle between the asymptotic direction of the light ray at the observer and the asymptotic direction at the emitter. The lens equations, Eqs. (\ref{WDLLensEq1}) and (\ref{WDLLensEq2}) have been derived up to and including contributions $\sim\varepsilon^2$ in \cite{SerenoLuca}.

Although that Kerr-Sen space-time is asymptotically flat the light ray is affected in a region near, but not too close to the black hole, such that $r_{H}\ll r_{\rm min}< r_{o/s}$. Following these relations we can extract from the geodesics the deflection angle as a function of the motion constants ${\cal{J}}$ and ${\cal{K}}$ and the black hole parameters $M$, $a$, $r_{\alpha}$. Considering source and observer with radial coordinates $r_{o}, r_{s}\rightarrow\infty$, we can neglect higher order terms $\sqrt{{\cal{J}}^2+{\cal{K}}}/r_{o}$ and $\sqrt{{\cal{J}}^2+{\cal{K}}}/r_{s}$ as well as their powers in the expressions of azimuthal and polar source coordinates, Eqs. (\ref{PolarLensEq}) and (\ref{AzimutalLensEq}). Taking the lensing geometry in consideration Eqs. (\ref{AngularSourceAngles1}), (\ref{AngularSourceAngles2}) and (\ref{AngularDiameterDistances2}) can be combined in such a way that for the deflection angle we can write the equations
\begin{eqnarray}\label{DeflectionAngle}
    \tan B_{1}(\mu_{s}|_{r_{o},r_{s}\rightarrow\infty}, \phi_{s}|_{r_{o},r_{s}\rightarrow\infty})&=&-\frac{D_{ds}}{D_{s}}\tan \hat{\alpha}_{1}, \\
    \tan B_{2}(\mu_{s}|_{r_{o},r_{s}\rightarrow\infty}, \phi_{s}|_{r_{o},r_{s}\rightarrow\infty})&=&-\frac{D_{ds}}{D_{s}}\tan \hat{\alpha}_{2},
\end{eqnarray}
which reduce to $\hat{\alpha}=\Delta\phi-\pi$, for an equatorial motion of the photons, $\mu_{s}=\mu_{o}=0$.
The bending of light ray by Kerr--Sen black hole leads to the following components of the deflection angle
\begin{eqnarray}\label{DA1}
  \hat{\alpha}_{1} &=& 4\frac{M}{b}\frac{b_{1}}{b}+\frac{15\pi}{4}\left(\frac{M}{b}\right)^2\frac{b_{1}}{b} -  \frac{\pi}{16}\frac{r_{\alpha}^2}{b^2}\frac{b_{1}}{b} + \frac{4(b_{1}^2-b_{2}^2)Ma\sqrt{1-\mu_{o}^2}}{b^4}  \nonumber \\
  &-& \frac{3\pi}{4}\frac{Mr_{\alpha}}{b^2}\frac{b_{1}}{b} + 64\left[ 1- \frac{1}{3}\left(\frac{b_{1}}{b}\right)^2 \right]\left(\frac{M}{b}\right)^3\frac{b_{1}}{b} - 4\left[ \frac{2(b_{2}^2-b_{1}^2)(1-\mu_{o}^2)}{b^2} + 1 \right] \frac{Ma^2}{b^3}\frac{b_{1}}{b} \nonumber  \\
  &-& \left[ \frac{5\pi\sqrt{1-\mu_{o}^2}(b_{2}^2-2b_{1}^2)}{b^2}-\frac{16b_{2}\mu_{o}}{b} \right]\frac{M^2a}{b^3} + \frac{\pi(b_{2}^2-2b_{1}^2)\sqrt{1-\mu_{o}^2}}{2b^2}\frac{Mar_{\alpha}}{b^3}  \nonumber \\
  &-& 16\frac{M^2r_{\alpha}}{b^3}\frac{b_{1}}{b} + \mathcal{O}(\epsilon^4),
\end{eqnarray}
\begin{eqnarray}\label{DA2}
  \hat{\alpha}_{2} &=& 4\frac{M}{b}\frac{b_{2}}{b}+\frac{15\pi}{4}\left(\frac{M}{b}\right)^2\frac{b_{2}}{b} -\frac{\pi}{16}\frac{r_{\alpha}^2}{b^2}\frac{b_{2}}{b} + \frac{8b_{1}b_{2}Ma\sqrt{1-\mu_{o}^2}}{b^4}  \\
  &-&\frac{3\pi}{4}\frac{M r_{\alpha}}{b^2}\frac{b_{2}}{b} + 64\left[ 1- \frac{1}{3}\left(\frac{b_{2}}{b}\right)^2 \right]\left(\frac{M}{b}\right)^3\frac{b_{2}}{b} \nonumber \\ &+& 2\left[ \left( \frac{b_{2}}{b} \right)^2\left(\left( \frac{b_{1}}{b_{2}} \right)^2-1\right)^2\mu_{o}^2+ \frac{2(b_{1}^2(3-2\mu_{o}^2)-b_{2}^2)}{b^2} \right]\frac{Ma^2}{b^3}\frac{b_{2}}{b} \nonumber \\
  &+& \left[ \frac{15\pi b_{1}b_{2}\sqrt{1-\mu_{o}^2}}{b^2}-\frac{16b_{1}\mu_{o}}{b} \right]\frac{M^2a}{b^3} - \frac{3\pi b_{1}b_{2} M r_{\alpha} a \sqrt{1-\mu_{o}^2}}{2b^6} -16\frac{M^2r_{\alpha}}{b^3}\frac{b_{2}}{b} + \mathcal{O}(\epsilon^4), \nonumber
\end{eqnarray}
where according to Eqs. (\ref{Observables3}) and (\ref{Observables4}) we have utilized with the components of the spherically symmetric impact parameter
\begin{eqnarray}
  b_{1} &=& -\frac{{\cal{J}}}{\sqrt{1-\mu_{o}^2}} \\
  b_{2} &=& -(-1)^k\sqrt{{\cal{K}}-{\cal{J}}^2\frac{\mu_{o}^2}{1-\mu_{o}^2}}.
\end{eqnarray}

For the case of the Kerr black hole ($r_{\alpha}=0$), Eqs. (\ref{DA1}) and (\ref{DA2}) reduce to the result in \cite{SerenoLuca}. Because of the presence of an charge, expressed through the parameter $r_{\alpha}$, the spin enters in the deflection angle coupled either to the black hole mass or with the mass and $r_{\alpha}$. In the static case ($a=0$) when the GMGHS black hole is realized the parameter $r_{\alpha}$ appears in the deflection angle independently as well coupled to the mass. The rotating black hole effects the motion of the photon most strongly when the whole trajectory is limited on the equatorial plane. Therefore it is the best case to study how the spin influences the deflection angle. Setting $b_{2}=0$, $\mu_{o}=0$ we have
\begin{eqnarray}\label{EquatorialDeflectionAngle}
  \hat{\alpha}_{1}&=&4\frac{M}{b}\frac{b_{1}}{b}+\frac{15\pi}{4}\left(\frac{M}{b}\right)^2\frac{b_{1}}{b}-\frac{\pi}{16}\frac{r_{\alpha}^2}{b^2}\frac{b_{1}}{b}
  +4\frac{Ma}{b^2}-\frac{3\pi}{4}\frac{Mr_{\alpha}}{b^2}\frac{b_{1}}{b}+\frac{128}{3}\left(\frac{M}{b}\right)^3\frac{b_{1}}{b}  \nonumber \\
  &+&4\frac{Ma^2}{b^3}\frac{b_{1}}{b}+10\pi\frac{M^2a}{b^3}-\pi\frac{Mar_{\alpha}}{b^3}-16\frac{M^2r_{\alpha}}{b^3}\frac{b_{1}}{b} + \mathcal{O}(\epsilon^4).
\end{eqnarray}
The bending of the light ray in the presence of black hole angular momentum depends on whether the motion of photons is in the direction of the spin, or in reverse direction. Compared with the GMGHS black hole case, the deflection angle is smaller for direct photons ($b_{1}<0$), and greater for the retrograde photons ($b_{1}>0$). Besides, the presence of an charge leads to appearance of non-zero contribution of order $\sim\mathcal{O}(\frac{Mar_{\alpha}}{b^3})$ in Eq. (\ref{EquatorialDeflectionAngle}), such that the deflection angle should be enhanced for prograde photons and reduced for retrograde photons. However, it is not the real picture. The contributions of higher orders are stronger and this effect is suppressed. For a given value of the angular momentum $a$, the growing of the parameter $r_{\alpha}$ results in a decreasing of the deflection angle without sense of the direction of winding of the photons with respect to the black hole rotation. When the angular momentum vanishes, compared with the Schwarzschild black hole case, the GMGHS black hole deflection angle is smaller due to the negative contribution of orders $\sim\mathcal{O}(\frac{r_{\alpha}}{b})$ and $\sim\mathcal{O}(\frac{Mr_{\alpha}}{b^2})$ as well as all their powers. For the case of the GMGHS black hole ($a=0$), Eq. (\ref{EquatorialDeflectionAngle}) reduces to the result in \cite{KeetonPetters} up to the given formal order in the expansion parameter.


\section{Critical Curves and Caustics}

The critical curves separate the regions in the lens plane where the Jacobian determinant $J$ of the lens map has opposite sign. For a point lens at these curves, the magnification factor of the images ${\cal{A}}$ diverges. Taking into account that the source emits isotropically, the observer will see the unlensed source $(r_{s}/r_{os})^2$ times smaller with respect to the observer positioned at the black hole \cite{BozzaLucaScarpettaSereno}. Then the Jacobian determinant reads
\begin{equation}\label{CritCurves}
       J=\left( \frac{r_{o}}{r_{os}} \right)^2\left( \frac{\partial\mu_{s}}{\partial\theta_{1}}\frac{\partial\phi_{s}}{\partial\theta_{2}}- \frac{\partial\mu_{s}}{\partial\theta_{2}}\frac{\partial\phi_{s}}{\partial\theta_{1}} \right).
\end{equation}
Representing the angular and the polar source positions $\{\mu_{s},\phi_{s}\}$ as a function of the angular coordinates of the images $\{\theta_{1},\theta_{2}\}$ up to the the post-Newtonian order the Jacobian is
\begin{equation}\label{Jacobian}
    J=1-\frac{1}{\theta_{(0)}^4}+\left(\left[1-\frac{r_{\alpha}}{5M}\right]\frac{15\pi(1
    -\theta_{(0)}^2)^2}{16(1+\theta_{(0)}^2)\theta_{(0)}^{5}}-\frac{4\theta_{1(0)}}{(1+\theta_{(0)}^2)\theta_{(0)}^4}\frac{a\sqrt{1-\mu_{o}^2}}{M}\right)\varepsilon+\mathcal{O}(\varepsilon^2),
\end{equation}
which reduces to the correction terms expected for Kerr black hole \cite{SerenoLuca} up to post-Newtonian order.

We look for a parametric solution up to the post-Newtonian order in the form
\begin{eqnarray}\label{Caust}
  \theta_{1}^{cr} &=&\theta_{\rm E}\cos\varphi\{1+\delta\theta_{\rm E}(\varphi)\varepsilon+\mathcal{O}(\varepsilon^2)\},  \nonumber \\
  \theta_{2}^{cr} &=&{\theta_{\rm E}}\sin\varphi\{1+\delta\theta_{\rm E}(\varphi)\varepsilon+\mathcal{O}(\varepsilon^2)\},
\end{eqnarray}
where $\varphi$ is an angle in a polar coordinate system taken in the lens plane with origin at the lens. In that system $\tan\varphi=\tan\theta_{1}{/}\tan\theta_{2}$. The first term of Eqs. (\ref{Caust}) gives the Schwarzschild black hole Einstein ring with radius $\theta_{\rm E}$. Solving Eq. (\ref{CritCurves}) we obtain the deviation coefficient
\begin{equation}\label{DC}
    \delta\theta_{\rm E}=\left[ 1-\frac{r_{\alpha}}{5M} \right]\frac{15\pi}{32}+\frac{a\sqrt{1-\mu_{o}^2}}{M}\cos{\varphi}.
\end{equation}
\begin{figure}
   \includegraphics[width=1\textwidth]{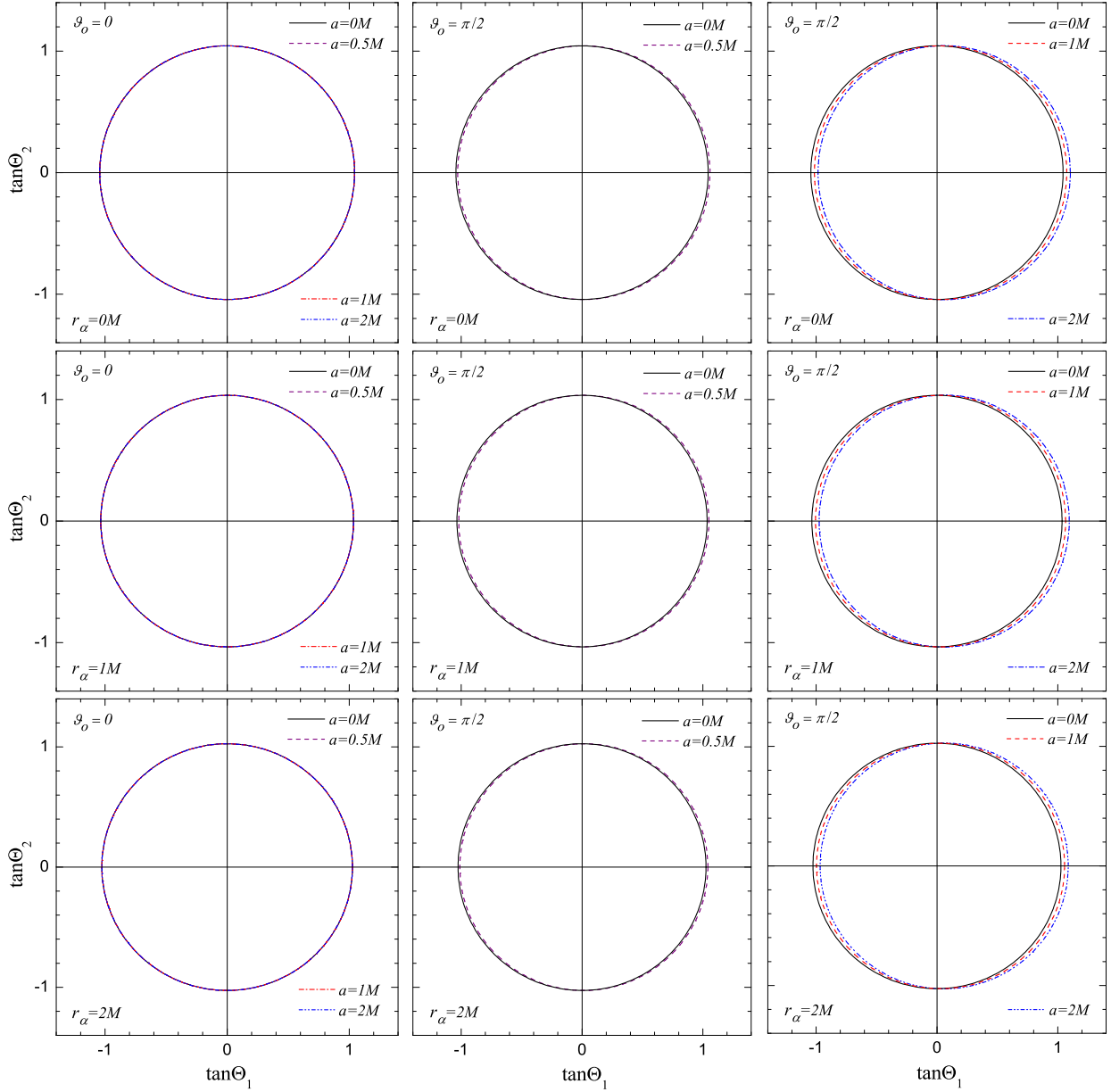}
   \caption{\small Critical curves in the plane $\{\tan\theta_{1},\tan\theta_{2}\}$ for an observer-lens position $r_{o}=7.62$ kpc and lens-source position $r_{s}=4.85\times10^{-5}$ pc. The Schwarzschild ($a=0M$, $r_{\alpha}=0M$), GMSHS ($a=0M$, $r_{\alpha}=1M$), Kerr ($a=0.5M$, $r_{\alpha}=0M$) and extremal Kerr black hole ($a=1M$, $r_{\alpha}=0M$) as well as GMGHS ($a=0M$, $r_{\alpha}=2M$) and Kerr ($a=2M$, $r_{\alpha}=0M$) naked singularity lenses are considered. We take notice also of the extremal Kerr--Sen black hole ($a=0.5M$, $r_{\alpha}=1M$) and Kerr-Sen naked singularities lenses ($a=1M$, $r_{\alpha}=1M$), ($a=2M$, $r_{\alpha}=1M$) and ($a=0.5M$, $r_{\alpha}=2M$) as well as ($a=1M$, $r_{\alpha}=2M$) and ($a=2M$, $r_{\alpha}=2M$). The observer is polar $\vartheta_{o}=0$ and equatorial $\vartheta_{o}=\pi/2$. Axis-lengths are in units of tangent of Einstein angle, $\theta_{E}\simeq157$ $\mu$arcsec.} \label{CCurves}
\end{figure}
For Kerr black hole lensing ($r_{\alpha}=0$), Eq. (\ref{DC}) reduces to the result by Sereno and De Luca \cite{SerenoLuca}. Setting the observer at position $\vartheta_{o}$ and fixing the values of the angular momentum $a$ and the parameter $r_{\alpha}$, we obtain a synonymous critical curve determined in the plane $\{\tan\theta_{1},\tan\theta_{2}\}$. The positions of the equatorial cross sections of the critical curves are shifted with respect to the static case by
\begin{equation}\label{CritShift}
   \tan\delta\theta^{cr}\simeq\frac{a\sqrt{1-\mu_{o}^2}}{r_{o}}+\mathcal{O}(\varepsilon^3).
\end{equation}

In order to calculate the caustics we have to find the corresponding source positions. Through Eqs. (\ref{AngularSourceAngles1}, \ref{AngularSourceAngles2}) and (\ref{Caust}), up to $\mathcal{O}(\varepsilon^3)$ order, the caustics are point-like and are positioned at
\begin{equation}\label{Caustic}
    \{ \tan B_{1}^{\rm cau}, \tan B_{2}^{\rm cau} \} \simeq \left\{ \frac{a\sqrt{1-\mu_{o}^2}}{r_{o}}+\mathcal{O}(\varepsilon^3), 0 \right\}.
\end{equation}
Our results for the critical curves shift, Eq. (\ref{CritShift}), and the caustic positions, Eq. (\ref{Caustic}), coincide with those found in \cite{SerenoLuca} up to the post-Newtonian order.

Critical curves are plotted in Fig. \ref{CCurves} for some values of the charge, the lens angular momentum and the observer's positions. We model the massive dark  object Sgr A* in the center of our Galaxy as a Gibbons--Maeda--Garfinkle--Horowitz--Strominger black hole and Kerr black hole as well as a Kerr--Sen black hole. Additionally, we also plotted the naked singularity cases, which are near to the black hole cases in the parametric space $\{a,r_{\alpha}\}$ according to the relations in Fig. \ref{BH-NS}. We assume a point source and set the lens between the source and the observer assuming that Sgr A* is located at a distance of $r_{o}=7.62$ kpc from Earth. Studying the influence of the lens parameters over the critical curves, as an illustration we made all graphics for a source positioned at a distance of $r_{s}=4.85\times10^{-5}$ pc from Sgr A*, so that it is outside the accretion disk. According to \cite{Eis} the lens has a mass $M=3.61 \times 10^6 M_{\odot}$. In this situation the expansion parameter $\varepsilon=0.029846679$.


\section{Image Positions}

Let us consider the solution of the perturbed lens equations, Eqs. (\ref{WDLLensEq1}) and (\ref{WDLLensEq2}). Assuming the series expansions, Eqs. (\ref{ImagePositions1}--\ref{ImagePositions3}), for the image positions, we can solve the lens equations term by term. At the first order in the deflection angle the lens equations reduce to the standard weak deflection lens equations describing the Schwarzschild lensing
\begin{eqnarray}
  \beta_{1} &=& \theta_{1(0)}\left( 1-\frac{1}{\theta_{(0)}^2} \right) \label{WDLE11} \\
  \beta_{2} &=& \theta_{2(0)}\left( 1-\frac{1}{\theta_{(0)}^2} \right) \label{WDLE12}
\end{eqnarray}
The square of the solution of the Newtonian order lens equation is $\theta_{(0)}^2=\theta_{1(0)}^2+\theta_{2(0)}^2$, where
\begin{eqnarray}
  \theta_{1(0)}^{\pm} &=& \frac{\beta_{1}}{2}\left(1\pm\sqrt{1+\frac{4}{\beta^2}}\right), \label{SWDLE11} \\
  \theta_{2(0)}^{\pm} &=& \frac{\beta_{2}}{2}\left(1\pm\sqrt{1+\frac{4}{\beta^2}}\right), \label{SWDLE12}
\end{eqnarray}
are the positive and the negative parity images, respectively. The square of the scaled angular source position is $\beta^{2}=\beta_{1}^2+\beta_{2}^2$. At this order the angular separation of the image with respect to the optical axis is
\begin{equation}\label{SWDLE1}
    \theta_{(0)}^{\pm}=\frac{\beta}{2}\left(1\pm\sqrt{1+\frac{4}{\beta^2}}\right).
\end{equation}
The positive parity image $\theta_{(0)}^{+}$ lies on the same side of the lens as the source and it is obtained for positive values of the scaled source position, $\beta>0$, while the negative parity image $\theta_{(0)}^{-}$ lies on the same side of the lens as the source and it is obtained for negative values of the scaled source position, $\beta<0$.

At the second order in the deflection angle the lens equations become
\begin{eqnarray}
  \left( 1-\frac{1}{\theta_{(0)}^2}\right)\theta_{1(1)}+\frac{2\theta_{1(0)}\theta_{2(0)}}{\theta_{(0)}^4}\theta_{2(1)}-
  \frac{15\pi}{16}\left(1-\frac{r_{\alpha}}{5M}\right)\frac{\theta_{1(0)}}{\theta_{(0)}^3}-\frac{\theta_{1(0)}^2-\theta_{2(0)}^2}{\theta_{(0)}^4}\frac{a\sqrt{1-\mu_{o}^2}}{M}=0, \label{WDLE21} \\
  \left( 1-\frac{1}{\theta_{(0)}^2}\right)\theta_{2(1)}+\frac{2\theta_{1(0)}\theta_{2(0)}}{\theta_{(0)}^4}\theta_{1(1)}-
  \frac{15\pi}{16}\left(1-\frac{r_{\alpha}}{5M}\right)\frac{\theta_{2(0)}}{\theta_{(0)}^3}-\frac{2\theta_{1(0)}\theta_{2(0)}}{\theta_{(0)}^4}\frac{a\sqrt{1-\mu_{o}^2}}{M}=0.  \label{WDLE22}
\end{eqnarray}
After deriving the post-Newtonian order correction of the lens equations, we find the second terms of the image positions
\begin{eqnarray}
  \theta_{1(1)} &=&
  \theta_{(1)}^{\rm GMGHS}\frac{\theta_{1(0)}}{\theta_{(0)}}+\frac{(1-\theta_{1(0)}^2+\theta_{2(0)}^2)}{1-\theta_{(0)}^4}\frac{a\sqrt{1-\mu_{o}^2}}{M}, \label{SWDLE21} \\
  \theta_{2(1)} &=&
  \theta_{(1)}^{\rm GMGHS}\frac{\theta_{2(0)}}{\theta_{(0)}}-\frac{2\theta_{1(0)}\theta_{2(0)}}{1-\theta_{(0)}^4}\frac{a\sqrt{1-\mu_{o}^2}}{M}, \label{SWDLE22}
\end{eqnarray}
and the angular separation of the image with respect to the optical axis
\begin{equation}\label{SWDLE2}
    \theta_{(1)}=\theta_{(1)}^{\rm GMGHS}+\frac{\theta_{1(0)}}{(1+\theta_{(0)}^2)\theta_{(0)}}\frac{a\sqrt{1-\mu_{o}^2}}{M},
\end{equation}
where
\begin{equation}\label{GMGHSImageCorection}
    \theta_{(1)}^{\rm GMGHS}=\left[1-\frac{r_{\alpha}}{5M}\right]\frac{15\pi}{16(1+\theta_{(0)}^2)}
\end{equation}
is the static and spherically symmetric first order correction of the Kerr--Sen image positions. Equations (\ref{SWDLE21}--\ref{SWDLE2}) reduce to the correction terms expected for rotating lenses \cite{SerenoLuca, WernerPetters} in the case of the Kerr black hole ($r_{\alpha}=0$). In the static case of the GMGHS black hole lens $(a=0)$, Eq. (\ref{SWDLE2}) reduces to the result found in \cite{KeetonPetters} up to the given formal order in the expansion parameter. The post-Newtonian corrections of the positive and negative parity images can be found since we already know $\theta_{1(0)}^{\pm}$, $\theta_{2(0)}^{\pm}$.

The angular size of GMGHS black hole image $\theta\simeq\theta_{\rm E}\{ \theta_{(0)}^{\pm}+\theta_{(1)}^{\rm GMGHS}\varepsilon+ \mathcal{O}(\varepsilon^2) \}$ is a solution of the most general lens equation for a static and spherically symmetric lens constructed in \cite{Bozza3}
\begin{equation}\label{SSSLE}
    r_{os}\sin{B}=r_{o}\sin{\theta}\cos{(\hat{\alpha}_{\rm GMGHS}-\theta)}-\sqrt{r_{s}^2-r_{o}^2\sin^2\theta}\sin{(\hat{\alpha}_{\rm GMGHS}-\theta)},
\end{equation}
where $\hat{\alpha}_{\rm GMGHS}$ is the Gibbons--Maeda--Garfinkle--Horowitz--Strominger black hole deflection angle, while $r_{os}=r_{o}\cos{B}+\sqrt{r_{s}^2+r_{o}^2\sin^2{B}}$ is the linear observer-source distance.

\begin{figure}
  \includegraphics[width=0.8\textwidth]{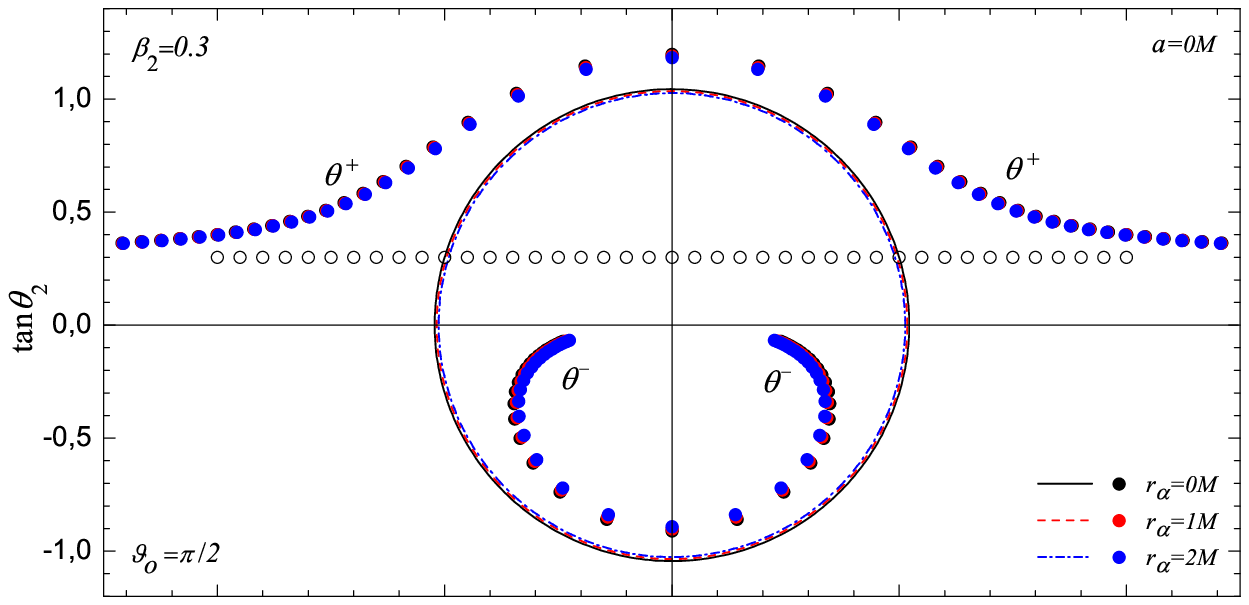}\\
  \includegraphics[width=0.8\textwidth]{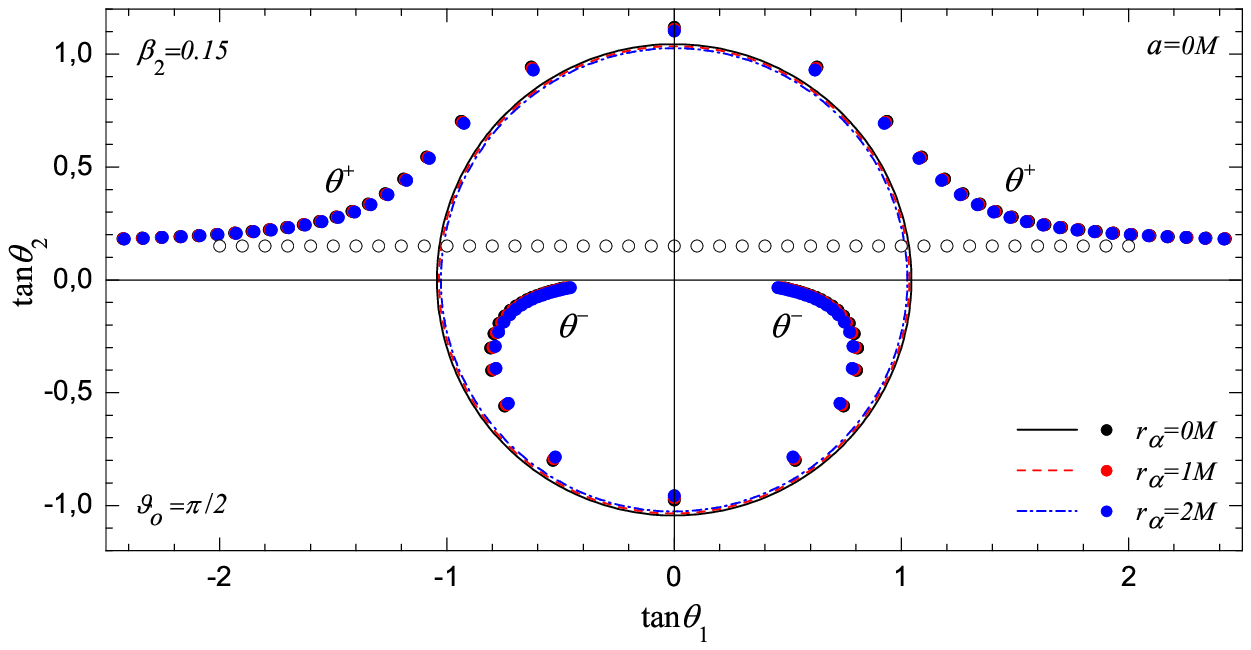}\\
  \caption{\small  The trajectory of the positive parity image $\theta^{+}$ and the shape of that of the negative parity image $\theta^{-}$ for a source (empty circlets) passing with scaled angular coordinates $\beta_{2}=0.15$ (below) and $\beta_{2}=0.3$ (above) over the equatorial plane. The corresponding critical curves are also plotted. The Schwarzschild black hole (filled black circlets, solid line) and GMGHS black hole (filled red circlets, dashed line) as well as GMGHS naked singularity lenses (filled blue circlets, dashed dotted line) are considered. The observer is equatorial $\vartheta_{o}=\pi/2$ at position $r_{o}=7.62$ kpc and the source is at position $r_{s}=4.85\times10^{-5}$ pc. Axis-lengths are in units of tangent of Einstein angle, $\theta_{E}\simeq157$ $\mu$arcsec.}\label{ImageTrajectories1}
\end{figure}
\begin{figure}
  \includegraphics[width=0.8\textwidth]{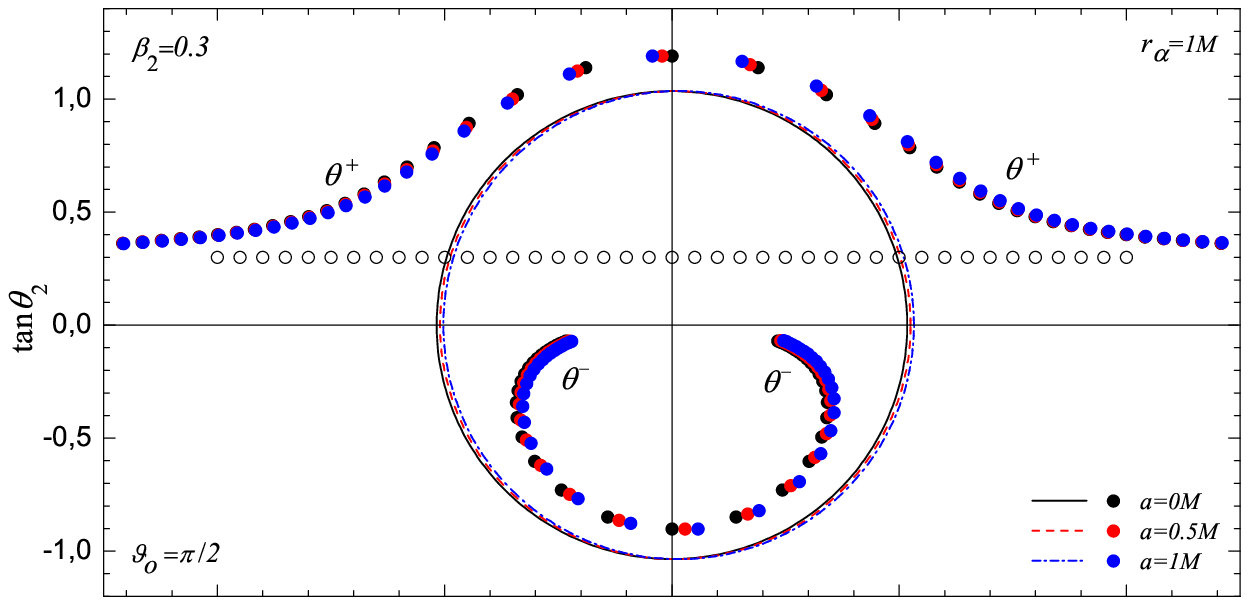}\\
  \includegraphics[width=0.8\textwidth]{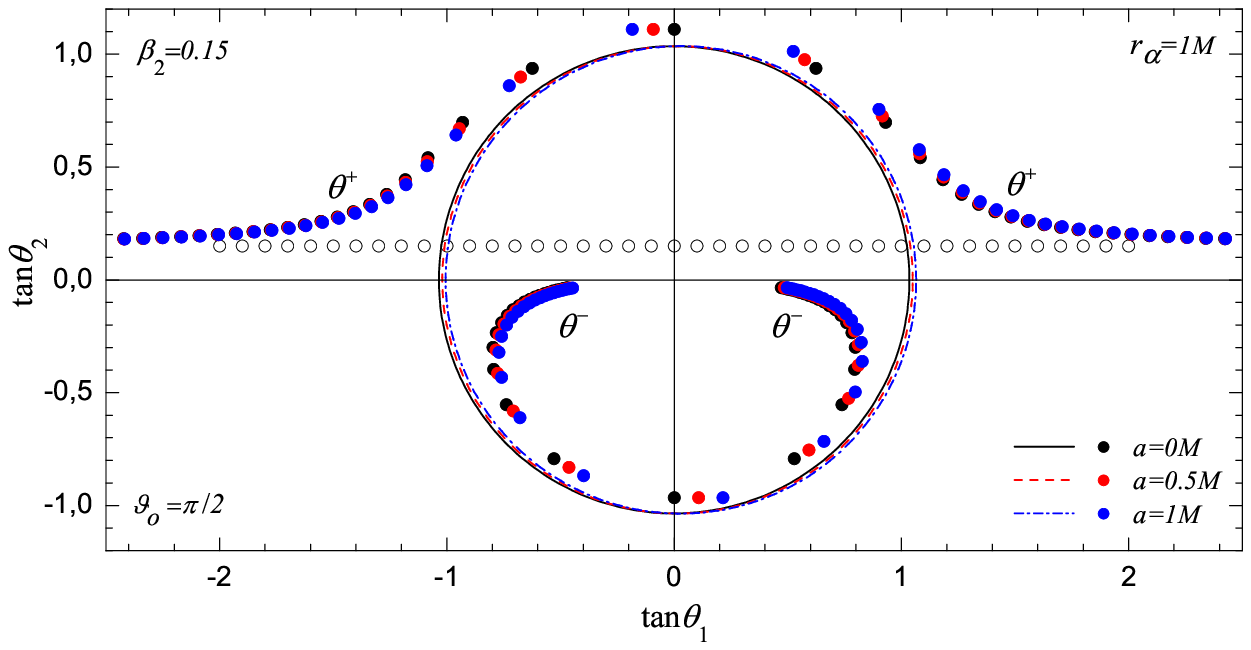}\\
  \caption{\small The trajectory of the positive parity image $\theta^{+}$ and the shape of that of the negative parity image $\theta^{-}$ for a source (empty circlets) passing with scaled angular coordinates $\beta_{2}=0.15$ (below) and $\beta_{2}=0.3$ (above) over the equatorial plane. The corresponding critical curves are also plotted. The GMGHS black hole (filled black circlets, solid line) and extremal Kerr--Sen black hole (filled red circlets, dashed line) as well as Kerr--Sen naked singularity lenses (filled blue circlets, dashed dotted line) are considered. The observer is equatorial $\vartheta_{o}=\pi/2$ at position $r_{o}=7.62$ kpc and the source is at position $r_{s}=4.85\times10^{-5}$ pc. Axis-lengths are in units of tangent of Einstein angle, $\theta_{E}\simeq157$ $\mu$arcsec.}\label{ImageTrajectories2}
\end{figure}

The results show that at this order there is a degeneracy between the Kerr-Sen black hole and the GMGHS black hole lenses, which seems to be displaced from the optical axis along the equatorial plane at $\{\theta_{1},\theta_{2}\}\simeq\theta_{\rm E}\left\{\frac{a\sqrt{1-\mu_{o}^2}}{M}\varepsilon, 0\right\}$.  When the lens angular momentum is oriented along the $Ox_{2}$ axis, photons which impact the lens with $b_{1}<0$ are positioned close to the center in contrast to the photons with $b_{1}>0$, which are placed further with respect to the center. The critical curves are shifted on the right hand side of the lens. When the lens is the static and spherically-symmetric GMGHS black hole the weak field images are symmetrically positioned with respect to the lens optical axis. If the source is detached from the equatorial plane, for every $\beta\neq\beta^{\rm cau}$ and $a>0$ the observer will see how the two weak field images will be counterclockwisely rotated, around the optical axis, with respect to the line connecting the static and spherically-symmetric images. When we move the source from the left hand side of the lens to the right hand side, such that the scaled source angular coordinate $\beta_{2}=const$, the two images follow the unperturbed ones. The effect of the parameter $r_{\alpha}$, which appears, consists in that, the radius of the critical curves as well as the angular separation of the image positions from the optical axis decrease with the increasing of the lens charge. When $r_{\alpha}=0$ the behaviour of the images is the same as in the Kerr black hole case, which has been described in \cite{SerenoLuca}. Considering a motion of light source in Fig. \ref{ImageTrajectories1} and Fig. \ref{ImageTrajectories2} we have plotted an image trajectories produced by the Schwarzschild, GMGHS and Kerr--Sen black hole lenses as well as GMGHS and Kerr-Sen naked singularity lenses for some values of the lens charge and angular momentum.

At the linear order in mass $M$, $r_{\alpha}$ and specific angular momentum $a$, the Kerr--Sen lensing is observationally equivalent to the gravitational lening by displaced GMSHS lens. If we take into account the nonlinear coupling between the physical parameters of the lens we can perform analytical calculations for the image positions up to and including terms $\sim\mathcal{O}(\varepsilon^2)$. As a result it has to be expected that the degeneracy between Kerr--Sen and GMSHS lenses will be broken (see Refs. \cite{AsadaKasai, SerenoLuca}). This is also seen from the expression of the deflection angle, which includes the squares of the angular momentum, $a^2$, and mass, $M^2$ in the terms of third order, $\mathcal{O}(\frac{M^2a}{b^3})$ and $\mathcal{O}(\frac{Ma^2}{b^3})$.


\section{Magnification}

According to the approximation of geometrical optics gravitational lensing causes a change in the cross section of a bundle of light rays, such that the surface brightness is conserved. Therefore, the ratio between the angular area element of the image in the celestial sky, $d\theta_{1}d\theta_{2}$, and the angular area element of the source in absence of the lens, $dB_{1}dB_{2}$, gives the signed image magnification $\cal{A}$, which is related to the Jacobian $J$ of the lens map
\begin{equation}\label{M1}
    {\cal{A}}=J^{-1}=\left( \frac{\partial B_{1}}{\partial\theta_{1}}\frac{\partial B_{2}}{\partial\theta_{2}}-\frac{\partial B_{1}}{\partial\theta_{2}}\frac{\partial B_{2}}{\partial\theta_{1}} \right)^{-1}.
\end{equation}

Using Eqs. (\ref{PolarLensEq}, \ref{AzimutalLensEq}) for the polar and the azimuthal source positions or the lens equations Eqs. (\ref{WDLLensEq1}, \ref{WDLLensEq2}) we obtain \begin{equation}\label{M2}
    {\cal{A}}=\frac{\theta_{(0)}^4}{\theta_{(0)}^4-1}-\left( \left[1-\frac{r_{\alpha}}{5M}\right]\frac{15\pi\theta_{(0)}^3}{16(1+\theta_{(0)}^2)^3}-\frac{4\theta_{(0)}^4\theta_{1(0)}}{(1-\theta_{(0)}^2)^2(1+\theta_{(0)}^2)^3}\frac{a\sqrt{1-\mu_{o}^2}}{M} \right)\varepsilon + \mathcal{O}(\varepsilon^2),
\end{equation}
which can be linked to the signed magnifications for both images. Equation (\ref{M2}) reduces to the magnification of the Kerr black hole images \cite{SerenoLuca}, when $r_{\alpha}=0$ and to the result found in \cite{KeetonPetters} for the static case of the GMGHS black hole lens $(a=0)$ up to the given formal order in the expansion parameter.

The individual magnifications of the positive and the negative parity image can be expressed in terms of the scaled angular source position in absence of the lens using Eqs. (\ref{SWDLE11}-\ref{SWDLE1}) up to post-Newtonian order. They are respectively
\begin{eqnarray}
  {\cal{A}}^{+} = \frac{(\beta+\sqrt{\beta^2+4})^4}{(\beta+\sqrt{\beta^2+4})^4-16}&-&\frac{1}{(4+\beta^2)^{3/2}} \nonumber  \\ &\times&\left(\left[1-\frac{r_{\alpha}}{5M}\right]\frac{15\pi}{16}-\frac{4\beta_{1}}{\beta^3}\frac{a\sqrt{1-\mu_{o}^2}}{M}\right)\varepsilon+\mathcal{O}(\varepsilon^2), \label{MuPlas} \\
  {\cal{A}}^{-} = \frac{(\beta-\sqrt{\beta^2+4})^4}{(\beta-\sqrt{\beta^2+4})^4-16}&-&\frac{1}{(4+\beta^2)^{3/2}} \nonumber \\ &\times&\left(\left[1-\frac{r_{\alpha}}{5M}\right]\frac{15\pi}{16}+\frac{4\beta_{1}}{\beta^3}\frac{a\sqrt{1-\mu_{o}^2}}{M}\right)\varepsilon+\mathcal{O}(\varepsilon^2). \label{MuMinus}
\end{eqnarray}
The contribution of the parameter $r_{\alpha}$ to the signed image magnifications is negative so that the Kerr-Sen weak field images will be slightly de-amplified.

Then the sum of the signed magnifications can be calculated and has the form
\begin{equation}\label{MagnSum}
    {\cal{A}}^{+}+{\cal{A}}^{-}=1-\frac{15\pi}{8(\beta^2+4)^{3/2}} \left[1-\frac{r_{\alpha}}{5M}\right]\varepsilon+\mathcal{O}(\varepsilon^2).
\end{equation}
The magnification invariant does not depend on the specific angular momentum $a$ and is equivalent to the result for GMGHS black hole lens to post-Newtonian order. At order $\mathcal{O}(\varepsilon)$ the deviation from the magnification invariant for the GMGHS lens and the Kerr--Sen lens are the same.

\begin{figure}
    \includegraphics[width=0.9\textwidth]{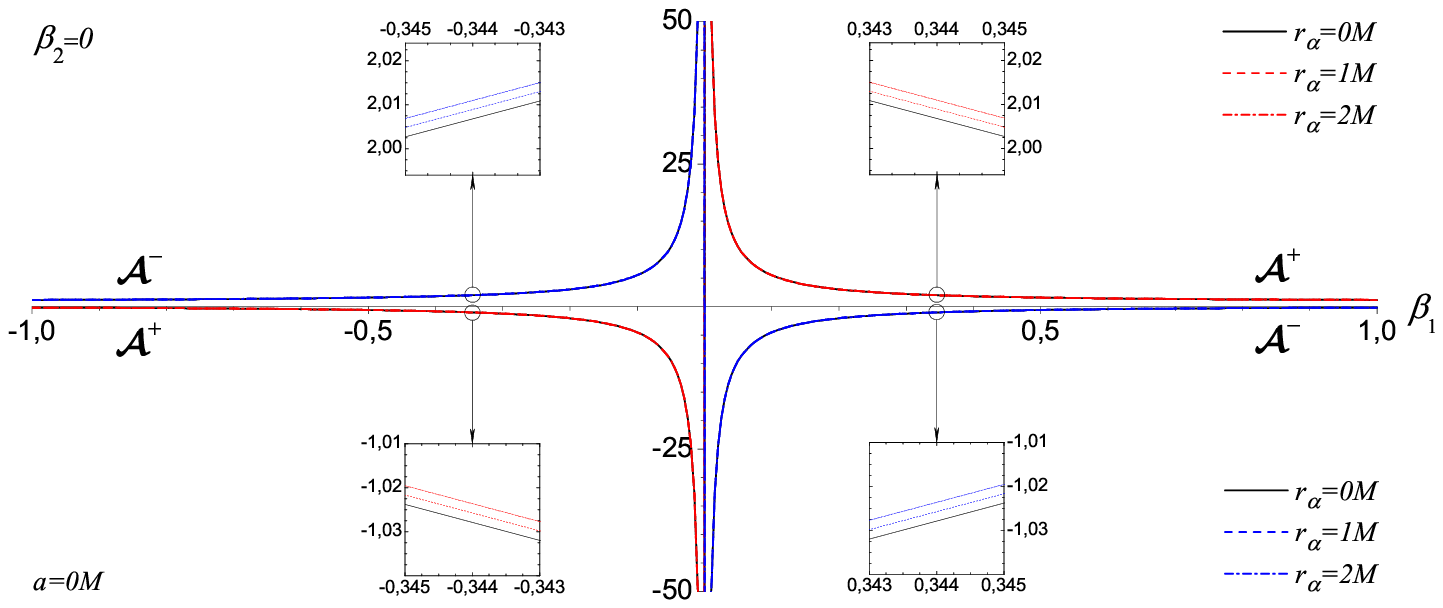} \\
    \vspace{0.5cm}
    \includegraphics[width=0.9\textwidth]{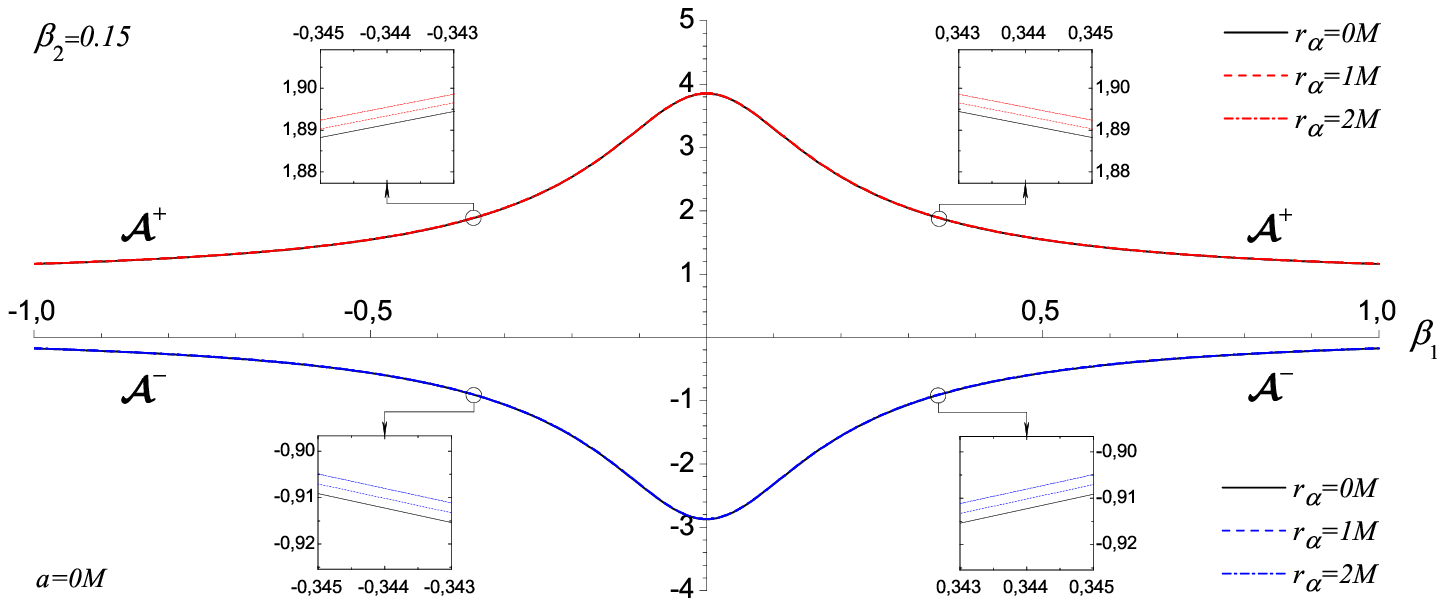} \\
    \vspace{0.5cm}
    \includegraphics[width=0.9\textwidth]{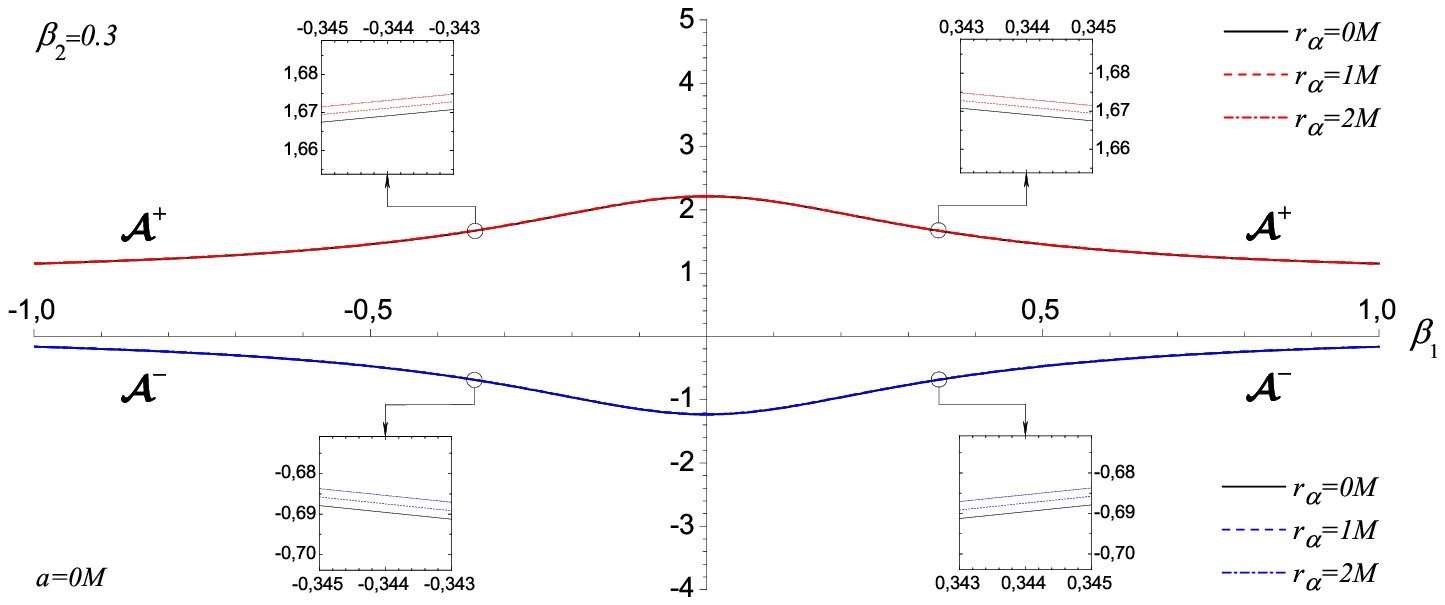}
    \caption{\small The magnification of the positive parity image ${\cal{A}}^{+}$ and the value of that of the negative parity image ${\cal{A}}^{-}$ as a function of the scaled angular coordinate $\beta_{1}$ of the source for different scaled angular coordinate $\beta_{2}$. The Schwarzschild black hole (solid line) and GMGHS black hole (dashed line) as well as GMGHS naked singularity lenses (dashed dotted line) are considered. The observer is equatorial $\vartheta_{o}=\pi/2$ at position $r_{o}=7.62$ kpc and the source is at position $r_{s}=4.85\times10^{-5}$ pc. The abscissa is in units of Einstein angle, $\theta_{\rm E}\simeq157$ $\mu$arcsec.} \label{SM1}
\end{figure}
\begin{figure}
    \includegraphics[width=0.9\textwidth]{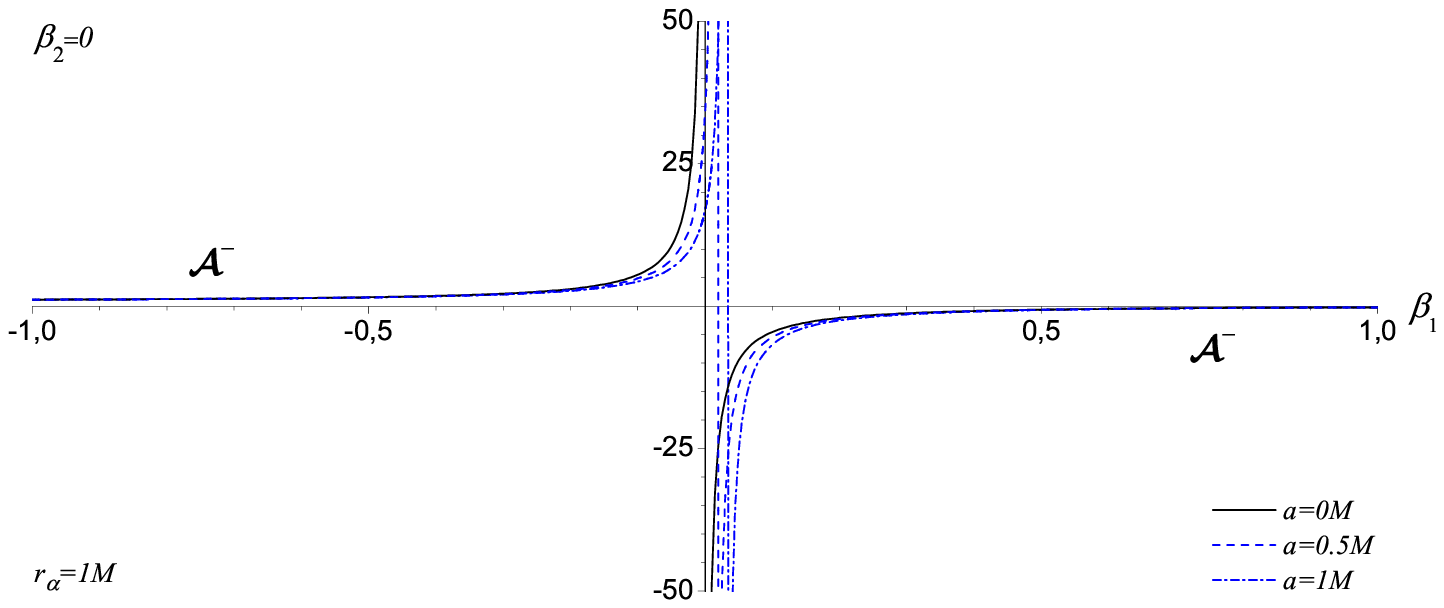} \\
    \vspace{1.0cm}
    \includegraphics[width=0.9\textwidth]{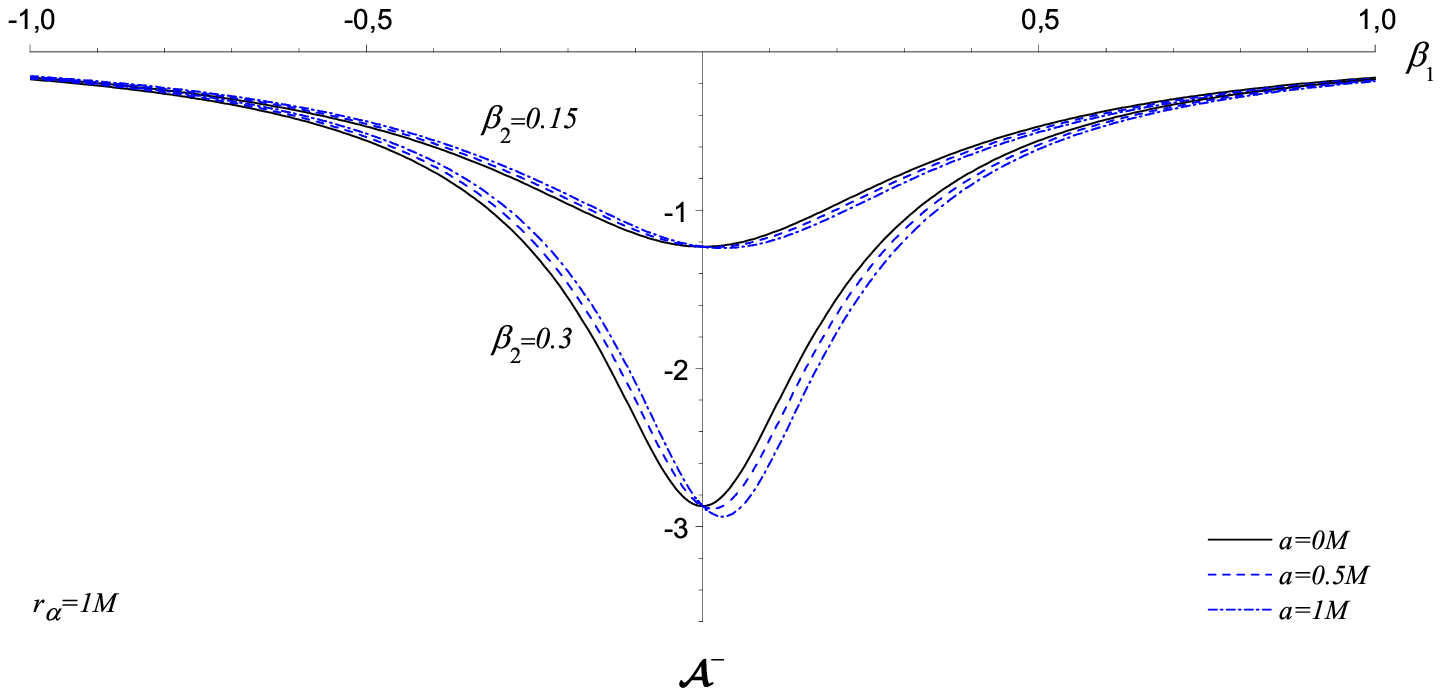} \\
    \caption{\small The magnification of the negative parity image ${\cal{A}}^{-}$ as a function of the scaled angular coordinate $\beta_{1}$ of the source for different scaled angular coordinate $\beta_{2}$. The GMGHS black hole (solid line) and extremal Kerr--Sen black hole (dashed line) as well as Kerr--Sen naked singularity (dash-dotted line) lenses are considered. The observer is equatorial $\vartheta_{o}=\pi/2$ at position $r_{o}=7.62$ kpc and the source is at position $r_{s}=4.85\times10^{-5}$ pc. The abscissa is in units of Einstein angle, $\theta_{\rm E}\simeq157$ $\mu$arcsec.} \label{SM2}
\end{figure}
\begin{figure}
    \includegraphics[width=0.9\textwidth]{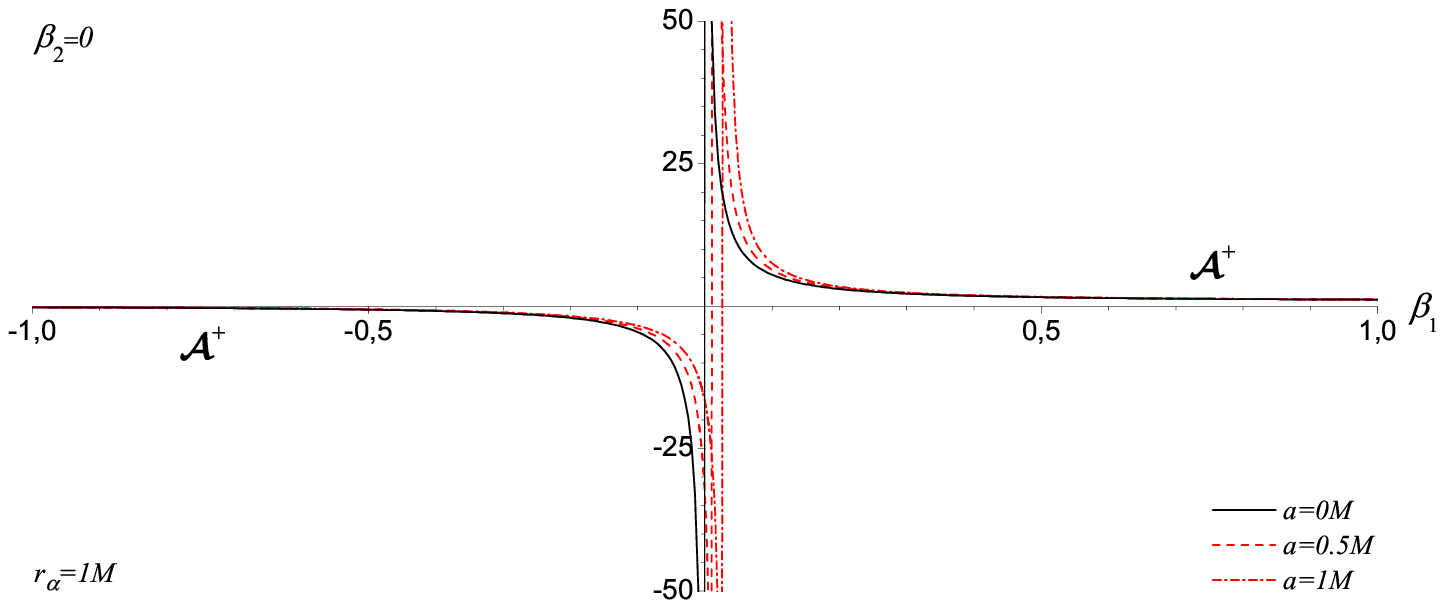} \\
    \vspace{1.0cm}
    \includegraphics[width=0.9\textwidth]{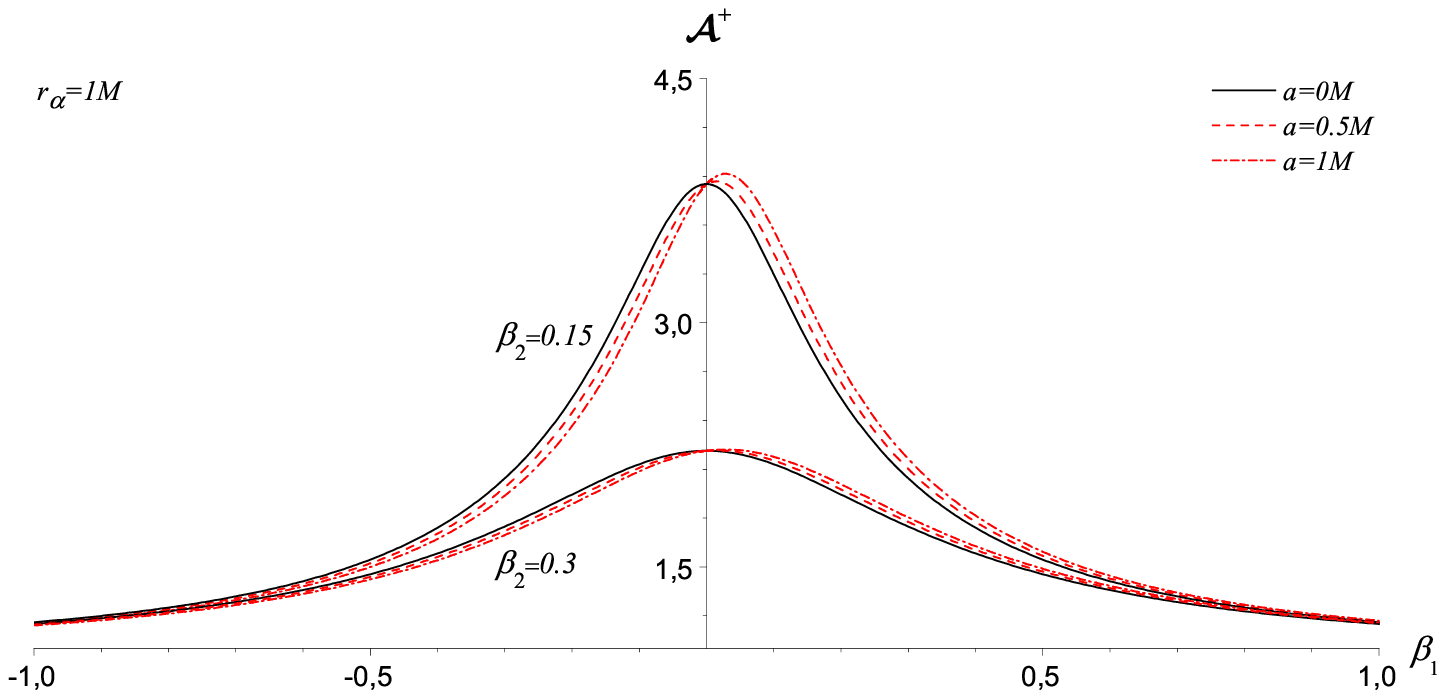} \\
    \caption{\small The magnification of the positive parity ${\cal{A}}^{+}$ as a function of the scaled angular coordinate $\beta_{1}$ of the source for different scaled angular coordinate $\beta_{2}$. The GMGHS black hole (solid line) and extremal Kerr--Sen black hole (dashed line) as well as Kerr--Sen naked singularity (dash-dotted line) lenses are considered. The observer is equatorial $\vartheta_{o}=\pi/2$ at position $r_{o}=7.62$ kpc and the source is at position $r_{s}=4.85\times10^{-5}$ pc. The abscissa is in units of Einstein angle, $\theta_{\rm E}\simeq157$ $\mu$arcsec.} \label{SM3}
\end{figure}

In the microlensing case, when the two weak field images are not resolved and are packed together, the main observables is the total magnification. Taking into account that the image parities give for the absolute magnifications $|{\cal{A}}^{+}|={\cal{A}}^{+}$ and $|{\cal{A}}^{-}|=-{\cal{A}}^{-}$, then the total absolute magnification for the Kerr--Sen space-time is
\begin{equation}\label{MangTot}
    {\cal{A}}_{\rm tot}=|{\cal{A}}^{+}|+|{\cal{A}}^{-}|=\frac{\beta^2+2}{\beta\sqrt{\beta^2+4}}+\frac{8\beta_{1}}{\beta^{3}(\beta^2+4)^{3/2}}\frac{a\sqrt{1-\mu_{o}^2}}{M}\varepsilon+\mathcal{O}(\varepsilon^2),
\end{equation}
and up to terms $\mathcal{O}(\varepsilon^2)$ does not differ from the result for Kerr lensing \cite{WernerPetters}. When the observer is on the rotational axis of the Kerr--Sen black hole or in the particular case of the circularly symmetric lens ($a=0$), the term $\mathcal{O}(\varepsilon)$ vanishes.

Let us discuss the behaviour of the image positions and their signed magnifications as regards the source motion as well as the change of the charge and the angular momentum of the black hole lens. In the vicinity of the caustics (critical curves) the power series (\ref{M2}) does not work properly and describes the image magnifications exactly for scaled source angular positions $\beta_{1}\neq\beta_{1}^{\rm cau}$ ($\theta_{1}\neq\theta^{cr}_{1}$). Therefore, in order to disentangle the influence of the charge and the lens angular momentum over the signed magnifications, in Figs. \ref{SM1}, \ref{SM2} and \ref{SM3} we have plotted the reciprocal values of the Jacobian corresponding to ${\cal{A}}^{+}$ and ${\cal{A}}^{-}$ according to the current estimates for the massive dark object in the center of our Galaxy \cite{Eis} as we assume that the observer is equatorial. For illustration, we have also plotted the naked singularity cases, which are near to the black holes in the parametric space $\{a,r_{\alpha}\}$ according to the relations in Fig. \ref{BH-NS}.

In the static case of the Gibbons--Maeda--Garfinkle--Horowitz--Strominger lens for all values of the scaled source angular positions $\beta_{1}$ and $\beta_{2}$ the magnification of the positive parity ${\cal{A}}^{+}$ and the value of that of the negative parity image ${\cal{A}}^{-}$ increases with the increase of the black hole charge. Let us set the source on the equatorial plane (\textit{i. e.} $\beta_{2}=0$) and move it to the optical axis. Then two weak field images will appear, one on each side of the lens. In the case $0<\beta_{1}$, the absolute values of signed image magnifications start to grow from ${\cal{A}}^{+}=1$ and ${\cal{A}}^{-}=0$ for the positive and negative parity images respectively when the source scaled angular coordinate is at infinity. Then the positive parity image has the source brightness, while the negative parity image is undistinguishable. In the opposite case $\beta_{1}<0$ the absolute values of signed image magnifications start to grow from ${\cal{A}}^{+}=0$ and ${\cal{A}}^{-}=1$ when the source scaled angular coordinate is at minus infinity. Then the positive parity image is undistinguishable, while the negative parity image has the source brightness. In the case when the source is on the optical axis ($(\beta_{1},\beta_{2})=(0,0)$) the signed magnifications diverge and infinitely bright Einstein ring appears. Removing the source from the optical axis (\textit{i. e.} $\beta_{2}\neq0$) and keeping $\beta_{1}=0$ we will see two weak field images situated respectively on each side of the optical axis in the directions perpendicular to the equatorial plane. Their magnifications decrease with the increase of $\beta_{2}$ and approach the limit values ${\cal{A}}^{+}=1$ and ${\cal{A}}^{-}=0$ for the positive and negative parity image respectively when the source scaled angular coordinate goes to infinity. In all of these cases the presence of the parameter $r_{\alpha}$ leads to an increment of the signed magnifications.

\begin{figure}
    \includegraphics[width=0.8\textwidth, height=0.25\textheight]{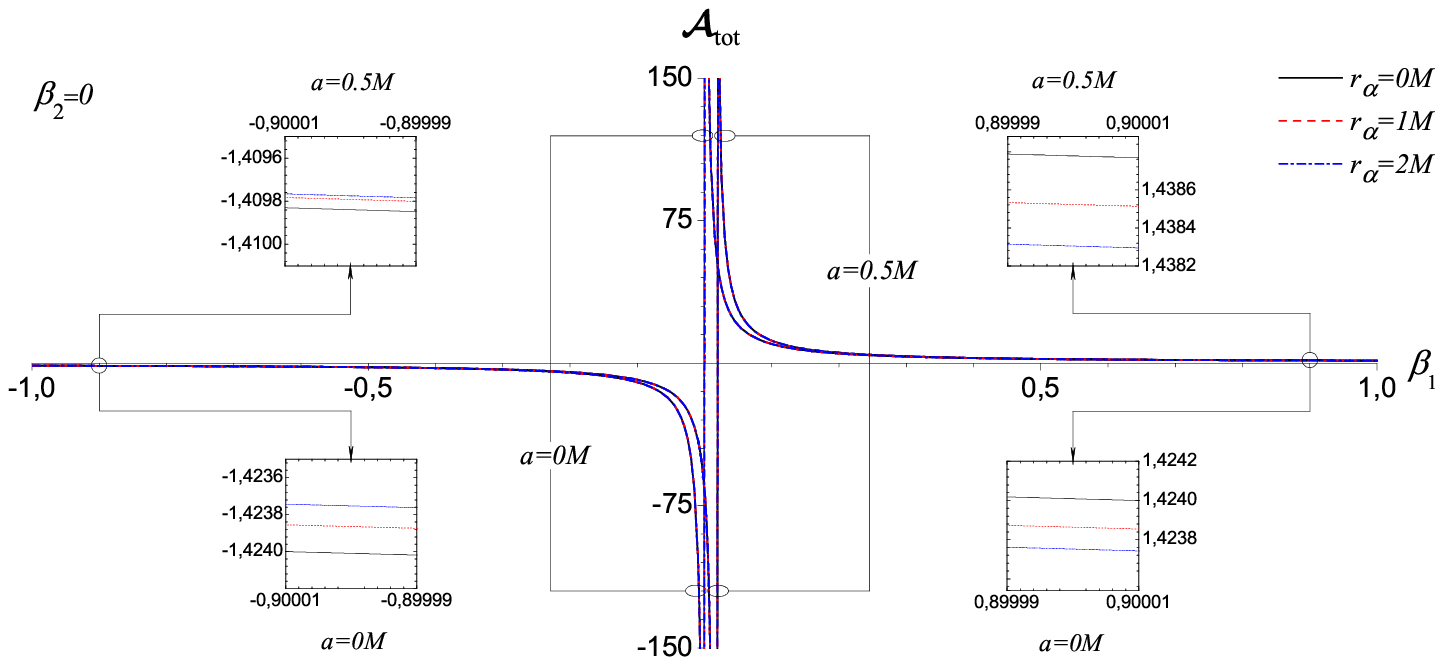} \\
    \vspace{0.5cm}
    \includegraphics[width=0.8\textwidth, height=0.25\textheight]{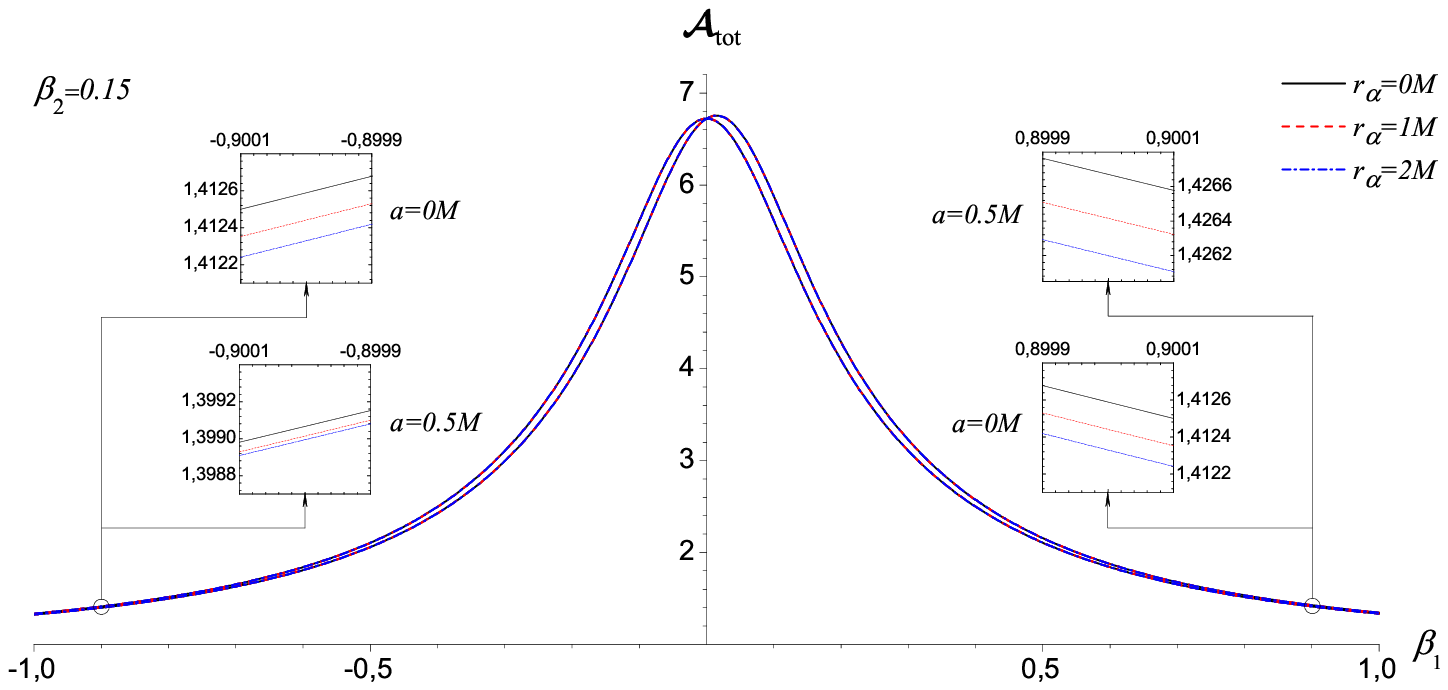} \\
    \vspace{0.5cm}
    \includegraphics[width=0.8\textwidth, height=0.25\textheight]{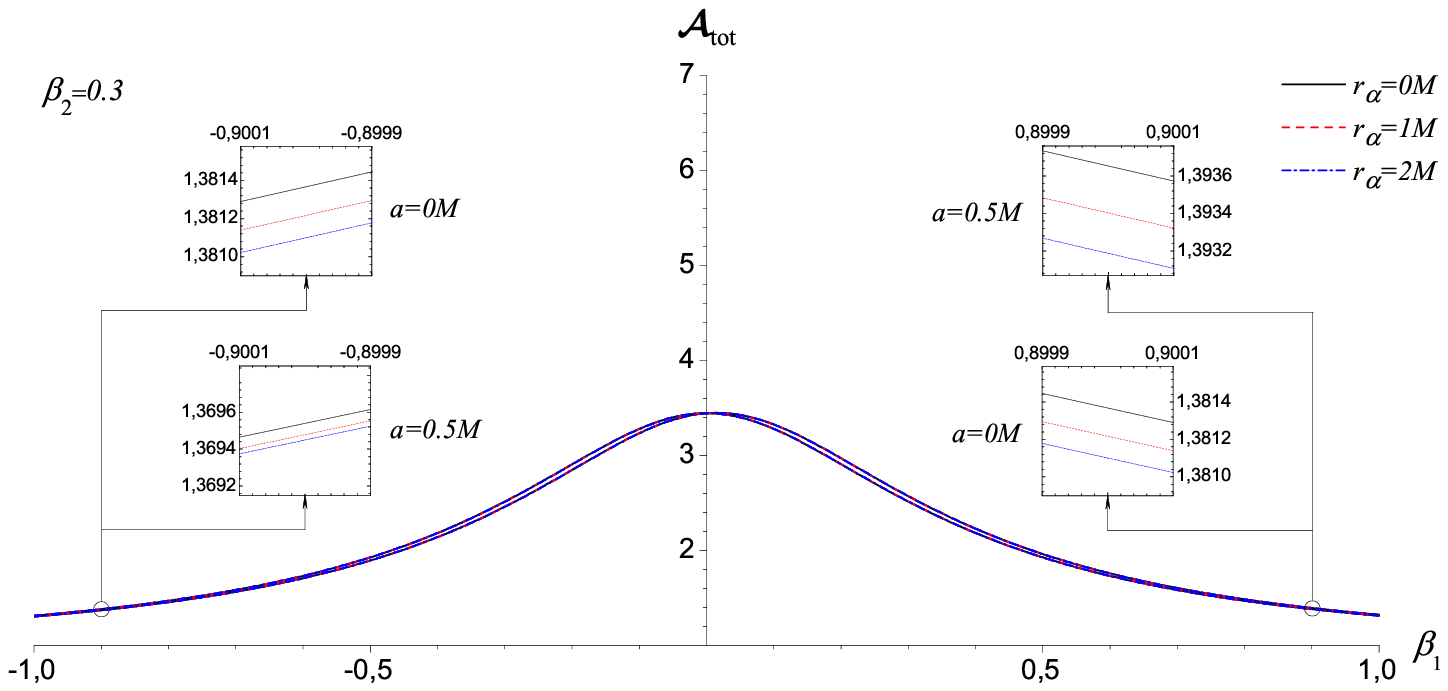}
    \caption{\small The total magnification ${\cal{A}}_{\rm tot}$ as a function of the scaled angular coordinate $\beta_{1}$ of the source for different scaled angular coordinate $\beta_{2}$. The Schwarzschild ($a=0M$ and $r_{\alpha}=0M$), GMGHS ($a=0M$ and $r_{\alpha}=1M$) and extremal Kerr--Sen ($a=0.5M$ and $r_{\alpha}=1M$) black hole lenses as well as GMGHS ($a=0M$ and $r_{\alpha}=2M$) and Kerr--Sen naked singularity lenses ($a=0.5M$ and $r_{\alpha}=2M$) are considered. The observer is equatorial $\vartheta_{O}=\pi/2$ at position $r_{o}=7.62$ kpc and the source is at position $r_{s}=4.85\times10^{-5}$ pc. The abscissa is in units of Einstein angle, $\theta_{\rm E}\simeq157$ $\mu$arcsec.} \label{TotMagn}
\end{figure}

When the lens is rotating and the source is equatorial ($\beta_{2}=0$) the angular momentum of the lens decreases the signed magnification of the negative parity image and increases the value of that of the positive parity image for every value of $\beta_{1}$ and charge. For the source position $\beta_{1}<\beta_{1}^{\rm cau}$ the negative parity image is outside the critical curve while the positive parity image is inside. Moving the source from infinity to the left hand side of the optical axis the negative parity image magnification starts to grow from ${\cal{A}}^{-}=1$ while the positive parity image magnification start to decrease from ${\cal{A}}^{+}=0$. Passing through the optical axis the source reaches the point-like caustic $(\beta_{1}^{\rm cau},0)=({a}/(\theta_{E}r_{o})+\mathcal{O}(\varepsilon^3),0)$, then the value of the signed image magnifications diverges and infinitely bright critical curves appear. Moving the equatorial source further towards the right hand side of the caustic point the negative signed magnification increases, while the positive signed magnification decreases with the increase of $\beta_{1}$ and approache respectively ${\cal{A}}^{-}=0$ and ${\cal{A}}{+}=1$ when $\beta_{1}\rightarrow\infty$. When we remove the source from the equatorial plane for every $\beta_{2}\neq0$ and $\beta_{1}<0$ the lens angular momentum leads to an increase in the negative parity magnification and to a decrease in the positive parity magnification in comparison to the magnification of the static case. When $\beta_{1}>0$ the negative parity magnification decreases, while the positive parity magnification increases. The signed magnifications coincide for a source in the point ($0,\beta_{2}$) where the static signed magnifications have a maximum. As a whole the non-equatorial source motion leads to occurrence of the positive parity image outside the critical curve and to rising of the negative parity image inside the critical curve. For fixed value of scaled angular coordinate $\beta_{1}$ and lens angular momentum $a$ as well as for all values of $\beta_{2}$ the values of the positive and the negative parity image magnifications increase with the increase of the lens charge.

Since the total magnification is the difference of the reciprocal values of the Jacobi determinant related to the positive and the negative parity images, we have plotted it in Fig. \ref{TotMagn} as a function of the source scaled angular coordinates $\beta_{1}$ and $\beta_{2}$ for different values of the lens angular momentum and the charge. ${\cal{A}}_{\rm tot}$ has a similar behavior as the positive parity magnification, with the difference that for fixed $a$ the total magnification decreases for $\beta_{1}>0$ and increases for $\beta_{1}<0$ with the increase of the charge. In the static case of the Gibbons--Maeda--Garfinkle--Horowitz--Strominger lens both the increment and the decrement of the total magnification have equal values for positive and negative source positions. This balance is violated in the rotational case of the Kerr-Sen lensing. Source standing on the right-hand side of the lens causes a decrement of the total magnification more than the total magnification increment provoked by a source on the left-hand side of the lens.


\section{Discussion and Conclusion}

In this paper we have discussed the features of gravitational lensing by a stationary, axially symmetric Kerr--Sen dilaton-axion black hole in the weak deflection limit. We derived the lens equations from the isotropic geodesic equations applying a perturbative formalism and represented the lensing quantities in series expansion by small parameter based on the weak-deflection angular Einstein ring radius. We performed our calculations up to and including third order terms and found that the non-zero contributions of the deflection angle are proportional to $M^3$, $a^2M$ and $aM^2$ as well as to $M^2r_{\alpha}$, $Mar_{\alpha}$ and $r_{\alpha}^2$. Modelling the massive compact object in the center of the galaxy as a Kerr--Sen black hole, we investigated the numerical values of the deflection angle of the light ray for the Schwarzschild, GMGHS, Kerr and Kerr--Sen black holes. Compared with the static and spherically symmetric case of the GMGHS black hole, the deflection angle in the rotating case is smaller for direct photons, and greater for the retrograde photons. As a whole, the effect of the charge, which appears, consists in that, the deflection angle decreases without sense of the direction of winding of the photons with respect to the black hole rotation. Based on the weak deflection limit analysis we found that up to the first order correction in the lens angular momentum, the Kerr--Sen lens is equivalent to a displaced GMGHS lens. We derived analytical descriptions for the two weak field images, the critical curves, the caustic points, the signed magnifications and the total magnification up to post-Newtonian order. Comparing the results to the corresponding quantities for the Kerr--Sen and Kerr black hole, we find that there are rotation and charge effects influent over the lensing observables.

The analytical results show that there are static post-Newtonian corrections to the two weak field image positions, to the signed magnification and to their sum as well as to the critical curves, which are function of the charge. The two perturbed images are counterclockwisely rotated, around the center, with respect to the line connected the static and spherically symmetric images. The angular separation of the image positions from the optical axis decreases with the increase of the charge, which causes a decreasing of the angular separation between the two images. This would open questions connected to their measuring in the microlensing cases when they are packed together and couldn't be resolved. All observable lensing quantities at the first order of the spin are a function of the projected over the lens plane angular momentum $a\sqrt{1-\mu_{o}}$, as expected. The critical curves are circle displaced along the equatorial direction, and it is shown that they decrease slightly for the Kerr--Sen lens with the increase of the charge. Up to the post-Newtonian order the caustics are point-like and do not depend of the charge.

In the static case of the Gibbons--Maeda--Garfinkle--Horowitz--Strominger lensing, the charge leads to a decrease of the signed magnifications for arbitrary source position. However, when the source is equatorial (\textit{i.e.} $\beta_{2}=0$), with the increase of the charge (or the parameter $r_{\alpha}$) the total magnification ${\cal{A}}_{\rm tot}$ decreases for a source on the right hand side of the optical axis (\textit{i.e.} $\beta_{1}>0$) and increases when the source is on the left hand side (\textit{i.e.} $\beta_{1}<0$). The non-equatorial source position leads only to a decrease of the total magnification for fixed $r_{\alpha}$. The critical curve are the weak deflection Einstein ring, which decreases almost undistinguishable with the increase of the metric parameter $r_{\alpha}$. The point-like caustic is positioned on the optical axis, as usual.

In the rotating case of the Kerr--Sen lensing, the point-like caustic drift away from the optical axis and some corrections with respect to the Gibbons--Maeda--Garfinkle--Horowitz--Strominger lenising appear. When the source is equatorial, for every values of $\beta_{1}$, the positive parity ${\cal{A}}^{+}$ and negative parity ${\cal{A}}^{-}$ magnifications increase or decrease respectively with the increase of $r_{\alpha}$ with respect to the signed magnifications in the static case. The non-equatorial source position leads to an increase of ${\cal{A}}^{+}$ for $\beta_{1}>0$ and to decrease of it for $\beta_{1}<0$ with respect to the static case. ${\cal{A}}^{-}$ has an opposite behavior with respect to ${\cal{A}}^{+}$. In this cases the charge increases the signed magnifications for all values of lens angular momentum $a$ and source positions $(\beta_{1}$, $\beta_{2})$. The total magnification reminds the behavior of positive parity magnification ${\cal{A}}^{+}$, with the difference that for fixed $a$ the total magnification decreases for $\beta_{1}>0$ and increases for $\beta_{1}<0$ with the increase of the charge.

For comparison with the case of the gravitational lensing by Kerr--Sen black hole in the strong deflection limit, we can refer to \cite{GyulchevYazadjiev} and to \cite{HiokiMiyamoto} for the gravitational capture of the photons by the dilaton-axion black hole and for the shadows observed at infinity.

All these results give us a reason to conclude that the Kerr--Sen black hole lensing in the weak deflection limit would be distinguishable with respect to the Kerr and Gibbons--Maeda--Garfinkle--Horowitz--Strominger black hole lensing. However, as a whole the dilaton-axion effect represses the values of the lensing observables for both the weak and the strong deflection limit lensing and therefore raises the question for using of imaging with hight sensitivity and resolution.

\begin{acknowledgments}
We wish to express our gratitude to M. Sereno for valuable comments and suggestions on the manuscript. This work was partially supported by the Bulgarian National Science Fund under Grants No. VUF-201/06 and No. DO 02-257, and the Sofia University Research Fund
No. 074/09.
\end{acknowledgments}

\appendix


\section{Radial integrals}

Here we briefly report the resolution strategy of the radial and angular integrals in the Kerr-Sen geodesic equations, Eqs. (\ref{IntEqMotion1}) and (\ref{IntEqMotion2}). Let us start with the radial integrals. In order to perform the integration we have to account for the sign convention in the geodesic equations as first. Namely, the integration over the whole trajectory of the photon must be made in such a way that all contributions from paths bounded by two consecutive inversion points to be sum up with the same sign \cite{Chandrasecar}. If the source and the observer are removed far away from the lens, the $r$ motion, $r_{s}\rightarrow r_{\rm min}\rightarrow r_{o}$ has only one inversion point so that we have to sum the contributions due to the approach and the departure of the photons. To solve analytically the integrals more easily it is recommended to change the variable $x=r_{\rm min}/r$. Then expanding the integrand as a Taylor series in $\epsilon_{m}$, $\epsilon_{a}$ and $\epsilon_{r_{\alpha}}$ we can perform the integration up to a given formal $n$-th order in $\epsilon=\epsilon_{m}^{i}\epsilon_{a}^{j}\epsilon_{r_{\alpha}}^{k}$, where $i+j+k\leq n$. In order to construct the lens equations it is necessity to evaluate the expanded primitive function in the extreme points $r_{s}$ and $r_{o}$ and hold back the terms proportional to $\sqrt{ {\cal{J}}^2+{\cal{K}} }/r_{s}$ and $\sqrt{ {\cal{J}}^2+{\cal{K}} }/r_{o}$ which are of order of $\epsilon$. With the above assumptions the integral of the left-hand side of Eq. (\ref{IntEqMotion1}) reads
\begin{eqnarray}
    \int_{r_{\rm min}}^{r_{s}}\frac{dr}{\sqrt{R(r)}}+\int_{r_{\rm min}}^{r_{o}}\frac{dr}{\sqrt{R(r)}} &\simeq& \frac{\pi}{\sqrt{{\cal{J}}^2+{\cal{K}}}} + \frac{4M}{{\cal{J}}^2 +{\cal{K}}} + \frac{15\pi M^2}{4({\cal{J}}^2 +{\cal{K}})^{3/2}} \nonumber \\
    &+& \frac{128M^3}{3({\cal{J}}^2+{\cal{K}})^{2}} + \frac{\pi(2{\cal{J}}^2-{\cal{K}})a^2}{4({\cal{J}}^2+{\cal{K}})^{5/2}} -  \frac{r_{\alpha}^2}{16({\cal{J}}^2+{\cal{K}})^{3/2}} \nonumber  \\
    &-& \frac{8{\cal{J}}Ma}{({\cal{J}}^2 +{\cal{K}})^{2}} - \frac{15\pi{\cal{J}}M^2a}{({\cal{J}}^2 +{\cal{K}})^{5/2}} + \frac{4(3{\cal{J}}^2-{\cal{K}})Ma^2}{({\cal{J}}^2 +{\cal{K}})^{3}}  \nonumber \\
    &-& \frac{3\pi M r_{\alpha}}{4({\cal{J}}^2 +{\cal{K}})^{3/2}} - \frac{16M^2r_{\alpha}}{({\cal{J}}^2 +{\cal{K}})^{2}} + \frac{3\pi{\cal{J}} M a r_{\alpha}}{2({\cal{J}}^2+{\cal{K}})^{5/2}} \nonumber \\
    &-& \frac{ {\cal{J}}^2+{\cal{K}} }{ 6r_{o}^3 } - \frac{ {\cal{J}}^2+{\cal{K}} }{ 6r_{s}^3 } - \frac{1}{r_{o}} - \frac{1}{r_{s}}.
\end{eqnarray}

Now we are in position to evaluate the second radial integral. Using Eq. (\ref{IntEqMotion1}) the radial part of Eq. (\ref{IntEqMotion2}) can be simplified. Then the integration is reduced to
\begin{eqnarray}
    \int_{r_{\rm min}}^{r_{s}} {\frac{a(2Mr-a{\cal{J}})}{\pm\Delta\sqrt{R(r)}}dr} + \int_{r_{\rm min}}^{r_{o}} \frac{a(2Mr-a{\cal{J}})}{\pm\Delta\sqrt{R(r)}}dr & \simeq &
    \frac{ 4Ma }{{\cal{J}}^2 +{\cal{K}}} - \frac{ {\cal{J}}\pi a^2}{2({\cal{J}}^2 +{\cal{K}})^{ 3/2 }} \nonumber \\
    &+& \frac{ 5\pi M^2 a }{({\cal{J}}^2 +{\cal{K}})^{ 3/2 }} - \frac{ 8{\cal{J}}Ma^2 }{({\cal{J}}^2 +{\cal{K}})^{ 2 }}   \nonumber \\
    &-& \frac{ \pi M a r_{\alpha} }{2({\cal{J}}^2 +{\cal{K}})^{ 3/2 }}.
\end{eqnarray}
The terms calculated at $r_{s}$ and $r_{o}$ are of order $\sim\epsilon^4$ and we have not consider them. An expressions for the calculated radial integrals up to the examined order in expansion parameters can be found in \cite{SerenoLuca} for the case of the Kerr black hole $(r_{\alpha}=0)$.


\section{Angular integrals}

The resolution of the angular integrals in the Kerr-Sen geodesic equations, Eqs. (\ref{IntEqMotion1}) and (\ref{IntEqMotion2}) lead to same results as these obtained for the case of the Kerr black hole \cite{SerenoLuca}. In consideration for completeness, let us represent a method for their solution. Introducing the useful variable $\mu=\cos\vartheta$ the angular integrals can be rewritten as
\begin{eqnarray}\label{AI1}
    I_{1}&=&\pm \int{\frac{ 1 }{ \sqrt{ \Theta_{\mu} } }d\mu}, \\
    I_{2}&=&\pm \int{\frac{ 1 }{ (1-\mu^2) \sqrt{ \Theta_{\mu} } }d\mu},
\end{eqnarray}
where
\begin{eqnarray}
   && \Theta_{\mu}=a^2(\mu_{-}^2+\mu^2)(\mu_{+}^2-\mu^2), \\
   && \mu_{\pm}^2=\frac{ \sqrt{ b_{{\cal{J}}{\cal{K}}}^2 + 4a^2{\cal{K}} } \pm b_{{\cal{J}}{\cal{K}}} }{ 2a^2 }, \label{Mu_Plus_Angular} \\
   && b_{{\cal{J}}{\cal{K}}}=a^2-{ {\cal{J}} }^2 - { {\cal{K}} }.
\end{eqnarray}
The turning point of the polar motion is a root of $\Theta_{\mu}=0$. Then the maximal and the minimal values that $\vartheta$ takes along the path can be determined from $\pm \mu_{+}$, with $-\mu_{+}$ corresponding to $\vartheta_{\rm max}$ and $\mu_{+}$ corresponding to $\vartheta_{\rm min}$. Using Eq. (\ref{Mu_Plus_Angular}) and expanding to the fourth order in $\epsilon=\epsilon_{a}$ we find
\begin{equation}\label{APCA}
    \mu_{+}=\sqrt{\frac{{\cal{K}}}{{\cal{J}}^2+{\cal{K}}}}\left[ 1+ \frac{ {\cal{J}}^2a^2 }{2({\cal{J}}^2+{\cal{K}})^2} + \frac{ {\cal{J}}^2(3{\cal{J}}^2-4{\cal{K}})a^4 }{8({\cal{J}}^2+{\cal{K}})^4} + \mathcal{O}(\epsilon^6) \right].
\end{equation}
Now we are in position to evaluate the integrals, which can be more easily done changing the variable to a new one $z=\mu/\mu_{+}$. The integrals become
\begin{eqnarray}
  J_{1} &=& \pm \int{\frac{ 1 }{ \sqrt{ \Theta_{z} } }dz}, \\
  J_{2} &=& \pm \int{\frac{ 1 }{ (1-\mu_{+}^2z^2) \sqrt{ \Theta_{z} } }dz},
\end{eqnarray}
where
\begin{equation}
    \Theta_{z}=a^2(\mu_{-}^2+\mu_{+}^2z^2)(1-z^2).
\end{equation}
Finally, after expanding the integrands to the second order in $\epsilon_{a}$ and performing the integration we get the primitive functions
\begin{eqnarray}\label{Angular_Int_1_z}
  PJ_{1}(z)  \simeq \frac{1}{\sqrt{{\cal{J}}^2+{\cal{K}}}}\left\{ \arcsin{z} + \left[ \frac{Kz\sqrt{1-z^2}}{ 4({\cal{J}}^2+{\cal{K}}) }  +  \frac{(2{\cal{J}}^2-{\cal{K}})}{ 4({\cal{J}}^2+{\cal{K}}) }\arcsin{z} \right]\frac{a^2}{{\cal{J}}^2+{\cal{K}}} \right\}
\end{eqnarray}
\begin{eqnarray}\label{Angular_Int_2_z}
  PJ_{2}(z)  \simeq \frac{{\cal{J}}}{|{\cal{J}}|}\arctan{ \frac{{\cal{J}}z}{\sqrt{{\cal{J}}^2+{\cal{K}}}\sqrt{1-z^2} } } - \left(   {\frac{{\cal{K}}z\sqrt{1-z^2}}{{\cal{J}}^2+{\cal{K}}(1-z^2)}-\arcsin{z}}\right) \nonumber \\
  \times \frac{ {\cal{J}}a^2 }{ 2({\cal{J}}^2+{\cal{K}})^{3/2} }
\end{eqnarray}

Since, the angular integrals appear with double signs, they must be performed according to the signs convention \cite{Chandrasecar}, namely all contributions from paths bounded by two consecutive inversion points must sum up with the same sign. The integration must start from the source position $z_{s}=\mu_{s}/\mu_{+}$ and must end at the observer position $z_{o}=\mu_{o}/\mu_{+}$. The emitted photon can take two direction at $z_{s}$. The former with initially growing of $z$ and the letter with initially decreasing of $z$. In the first case the first part of the angular integral covers the interval $[z_{s},1]$, which corresponds to motion of a photon from the source to the turning point $\mu_{+}$. The second part of the angular integral covers the interval $[z_{o},1]$, which corresponds to motion from the turning point $\mu_{+}$ to the observer. In the second case the first part of the angular integral covers the interval $[-1,z_{s}]$, which corresponds to motion of a photon from the source to the turning point $-\mu_{+}$. The second part of the angular integral covers the interval $[-1,z_{o}]$, which corresponds to motion from the turning point $-\mu_{+}$ to the observer. Then using the sign convention and the property of the primitive functions $PJ_{i}$ that $PJ_{i}(z)=-PJ_{i}(-z)$, we can express the angular integrals in the form
\begin{eqnarray}
  I_{i}(z) &=& 2PJ_{i}(1)+(-1)^{k}[PJ_{i}(z_{s})+PJ_{i}(z_{o})],
\end{eqnarray}
where everywhere $i=1,2$. $k$ is an integer number defined to be even if the photon gets to the observer from below to above (after having attained $\vartheta_{\rm max}$) and odd if the photon hits the observer from above to below (after having attained $\vartheta_{\rm min}$).

In terms of the variable $\mu$ the angular integrals read
\begin{eqnarray}
    PI_{1}(\mu) & \simeq & \frac{1}{ \sqrt{{\cal{J}}^2+{\cal{K}}} } \left\{ \arcsin{\mu_{\sigma}}
    - \left[ \frac{ 2{\cal{J}}^2-{\cal{K}}(1-\mu_{\sigma}^2) }{  4({\cal{J}}^2+{\cal{K}}) }  \right. \right. \nonumber \\
    &\times& \left. \left. \frac{\mu_{\sigma}}{\sqrt{1-\mu_{\sigma}^2}} + \frac{ {\cal{K}}-2{\cal{J}}^2 }{ 4({\cal{J}}^2+{\cal{K}}) }\arcsin{\mu_{\sigma}}  \right] \frac{a^2}{{\cal{J}}^2+{\cal{K}}} \right\}
\end{eqnarray}
\begin{eqnarray}
    PI_{2}(\mu) &\simeq& \frac{{\cal{J}}}{|{\cal{J}}|}\arctan{ \frac{{\cal{J}}\mu_{\sigma}}{ \sqrt{ {\cal{J}}^2+{\cal{K}} }\sqrt{1-\mu_{\sigma}} } } \nonumber \\
    &-&\left( \frac{\mu_{\sigma}}{\sqrt{1-\mu_{\sigma}^2}}-\arcsin{\mu_{\sigma}} \right)\frac{ {\cal{J}}a^2 }{ 2({\cal{J}}^2+{\cal{K}})^{3/2} },
\end{eqnarray}
where $\mu_{\sigma}=\mu\sqrt{({\cal{J}}^2+{\cal{K}})/{\cal{K}}}$. An exact analytical solution of the angular integrals in terms of elliptic integrals can be found in \cite{SerenoLuca}. Exploiting Eqs. (\ref{Angular_Int_1_z}) and (\ref{Angular_Int_2_z}) in the turning point $\mu_{+}$ ($z=1$) the primitive functions reduce to
\begin{equation}
    PI_{1}(\mu_{+})=\frac{\pi}{2\sqrt{{\cal{J}}^2+{\cal{K}}}}+\frac{\pi(2{{\cal{J}}^2}-{\cal{K}})a^2}{8({\cal{J}}^2+{\cal{K}})^{5/2}}.
\end{equation}

\begin{equation}
    PI_{2}(\mu_{+})=\frac{{\cal{J}}\pi}{2|{\cal{J}}|}+\frac{{\cal{J}}\pi a^2}{4({\cal{J}}^2+{\cal{K}})^{3/2}}.
\end{equation}

Finally, in terms of $\mu$ the path following from the angular integrals can be express in the compact form
\begin{eqnarray}
  I_{i}(\mu) &=& 2PI_{i}(\mu_{+})+(-1)^{k}[PI_{i}(\mu_{s})+PI_{i}(\mu_{o})],
\end{eqnarray}
with $i=1,2$.



\end{document}